\documentclass[%
 reprint,
 amsmath,amssymb,
 aps,
prf
]{revtex4-1}

\usepackage{graphicx}
\usepackage{dcolumn}
\usepackage{bm}
\usepackage{color}

\newcommand{\partiel}[2]{\frac{\partial #1}{\partial #2}} 

\begin{document}

\preprint{APS/123-QED}

\title{Effect of viscosity ratio on the self-sustained instabilities in planar immiscible jets}

\author{Outi Tammisola}
\altaffiliation[Also at ]{Faculty of Engineering, The University of Nottingham.\\
University Park, NG72RD, UK.}
\author{Jean-Christophe Loiseau}%
\author{Luca Brandt}%
\affiliation{
KTH Mechanics, KTH Royal Institute of Technology\\
Osquars backe 18, SE-10044 Stockholm, Sweden.}%

\date{\today}

\begin{abstract}
Previous studies have shown that intermediate magnitude of surface tension has a counterintuitive destabilizing effect on two-phase planar jets. In the present study, the transition process in confined two-dimensional jets of two fluids with varying viscosity ratio is investigated using direct numerical simulations (DNS). The outer fluid co-flow velocity is 17$\%$ of that of the central jet. Neutral curves for the appearance of persistent oscillations are found by recording the norm of the velocity residuals in DNS for over $1000$ nondimensional time units, or until the signal has reached a constant level in a logarithmic scale - either a converged steady state, or a "statistically steady" oscillatory state.  Oscillatory final states are found for all viscosity ratios ranging from $10^{-1}$ to $10$. For uniform viscosity ($m=1$), the first bifurcation is through a surface tension-driven global instability. On the other hand, for low viscosity of the outer fluid, there is a mode competition between a steady asymmetric Coanda-type attachment mode and the surface tension-induced mode. At moderate surface tension, the first bifurcation is through the Coanda-type attachment which eventually triggers time-dependent convective bursts. At high surface tension, the first bifurcation is through the surface tension-dominated mode. For high viscosity of the outer fluid, persistent oscillations appear due to a strong convective instability, although it is shown that absolute instability may be possible at even higher viscosity ratios. Finally, we show that the jet is still convectively and absolutely unstable far from the inlet when the shear profile is nearly constant. Comparing this situation to a parallel Couette flow (without inflection points), we show that in both flows, a hidden interfacial mode brought out by surface tension becomes temporally and absolutely unstable in an intermediate Weber and Reynolds regime. By an energy analysis of the Couette flow case, we show that surface tension, although dissipative, can induce a velocity field near the interface which extracts energy from the flow through a viscous mechanism. This study highlights the rich dynamics of immiscible planar uniform-density jets, where different self-sustained and convective mechanisms compete and the nature of the instability depends on the exact parameter values.
\begin{description}
\item[Usage]
  This is a preprint submitted to Physical Review F.
\item[PACS numbers]
  May be entered using the \verb+\pacs{#1}+ command.
\end{description}

\end{abstract}

\pacs{Valid PACS appear here}
\maketitle


\section{\label{sec:level1}Introduction}

Two-phase flows are encountered in numerous industrial applications, such as oil and gas transport, the atomization of jets in fuel injectors or even in microfluidics. Over the past decades, the understanding of the initial stage of transition to turbulence in such two-phase flows has essentially relied on the use of {\it local stability theory}. This local approach, based on the parallel flow assumption, allows one to investigate the linear stability of uni-directional base flows toward infinitesimal perturbations having a given streamwise and/or spanwise periodicity. Using this local ansatz, Boomkamp \& Miesen~\cite{Boomkamp_IJMF_1996} have proposed a classification of the different linear instability mechanisms existing in interfacial flows based on a careful inspection of the perturbation energy budget. Their review indicated that, in many of the two-phase flow situations investigated until then, the dominant instability mechanism results from a viscosity stratification which leads to net work being done by the perturbation velocity and stress at the interface separating the two phases. As for single phase flows, shear-driven linear instabilities (such as Tollmien-Schlichting waves) are also of importance and can even compete with the viscosity stratification mechanism as was shown by Yecko, Zaleski \& Fullana~\cite{yecko2002} for two-phase mixing layers. On the other hand, viscosity stratification may also invoke other instability mechanisms such as the short-wave instability. 

One mechanism by which the viscosity stratification causes instability in confined shear flows is the long-wave Yih mechanism~\cite{yih,Yiantsios_PoF_1988}. The seminal work of Yih found that Couette and Poiseuille flows become unstable for all Reynolds numbers if the outer fluid is more viscous \cite{yih}. Later, Hooper \& Boyd~\cite{hooperboyd1983} showed that Couette flow of two fluids in the absence of surface tension is always unstable to short waves. The mechanism for the short-wave instability due to viscosity stratification was analysed by Hinch~\cite{hinch1984}. 

These studies were based on an initally parallel base flow and rely on a local temporal stability approach to explain the initial stage of transition. In such a local temporal framework, surface tension is often either negligible or has a stabilizing effect on the instability (see \textit{e.g.} the classification by Boomkamp \& Miesen~\cite{Boomkamp_IJMF_1996}). However, when investigating the local absolute instability properties of a top-hat wake profile, Rees \& Juniper \cite{rees09} observed that surface tension increases the absolute growth rate of the instability in an inviscid model problem. 

Absolute instabilities can be related to what is known as {\it global instability modes} in non-parallel flows. Investigating the stability of such non-parallel flows (what is now known as {\it global linear stability}) has proven helpful in numerous single-phase flow situations to get a better understanding of the underlying physics. Unfortunately, probably because of its computational complexity, such a global approach to linear instability is still scarcely used to investigate strongly non-parallel two-phase flows. To our knowledge, Tammisola, Lundell \& S{\"o}derberg~\cite{VakarWe,StableWe} were first to use the global approach on two-fluid flows by solving the linearized Navier-Stokes equations numerically in both phases (and not treating either as inviscid in which case analytical solutions could have been found). In~\cite{StableWe}, the global instability of two-phase confined co-flowing jets and wakes with constant density and viscosity was investigated. Intermediate values of surface tension were found to cause global instability in jet flows which were robustly globally stable otherwise. For wakes~\cite{VakarWe,StableWe}, intermediate surface tension gives rise to global modes with considerably higher growth rates than the von K\`arm\`an mode. In both cases, strong enough surface tension eventually stabilises the global instability modes. Biancofiore, Gallaire \& Heifetz~\cite{biancofiore_surftens_pof} provided a physical explanation for the counterintuitive destabilization of wake flows by intermediate surface tension. The system was modelled as a broken-line shear layer, where counterpropagating Rossby waves formed at the vorticity discontinuities and capillary waves at the interface. By considering the resulting wave interaction they deduced that intermediate surface tension could cause local temporal and absolute instability.

The global linear predictions~\cite{VakarWe} on the wake flows were partly confirmed by Biancofiore et al.~\cite{Biancofiore_FDR_2014} using direct numerical simulations (DNS). While their calculations reveal relatively good agreements regarding the promotion of wake instability at intermediate surface tension, they did not observe at all the varicose instability modes for wakes predicted by global linear instability~\cite{StableWe}. A possible explanation given by the authors~\cite{Biancofiore_FDR_2014} was that the base flows in~\cite{StableWe} were computed in the absence of surface tension. Also, the observed wake instability due to surface tension saturated nonlinearly to such a low amplitude that the interface remained flat. Hence, the large effects on both jets and wakes indicated by linear global analysis still remained to be confirmed in nonlinear simulations. All that the DNS~\cite{Biancofiore_FDR_2014} seemed to show was that the surface tension merely altered the von Karman instability of wakes to another, very weak, instability mode. 

Moreover, the above-mentioned studies assumed the same density and viscosity of the two fluids. This assumption hardly holds for a generic two-fluid flow, and raises the question how density and viscosity ratio might alter the global instability behavior. A priori, the influence of viscosity ratio is hard to predict. The viscosity ratio could act by simply changing the effective Reynolds number of the two-fluid flow, thus changing the critical Reynolds number accordingly. However, there could be an interplay of different instability mechanisms. The study of Tammisola {\it et al.}~\cite{StableWe} identified surface tension and inflow shear as the dominant parameters for viscosity ratio $m=1$. However, as shown in numerous local stability analyses, viscosity stratification often drives the local instability properties of parallel flows through either the Yih mechanism~\cite{yih} or the short-wave instability mechanism~\cite{hooperboyd1983}. Furthermore, viscosity may affect the instability by changing the spatial development of the two-dimensional base flow, such as, the presence of recirculation regions. 

The present study takes on from the previous studies on co-flow jets and wakes, and has two main aims: (i) confirm the surprising destabilization of jet flows in nonlinear simulations, and (ii) investigate how viscosity ratio affects the presence of self-sustained oscillations (global instability). To this end, direct numerical simulations (DNS) are performed using our in-house DNS solver OILS (Optimised Interfacial Level Set), derived from the Two-Phase Level Set (TPLS) open-source code \cite{TPLS-web-page}, which was successfully used to study the nonlinear development of the Yih instability recently~ \cite{ONaraigh_JFM_2014}. 

\section{Problem statement}

\begin{figure}
\begin{center}
   \includegraphics[width=\columnwidth]{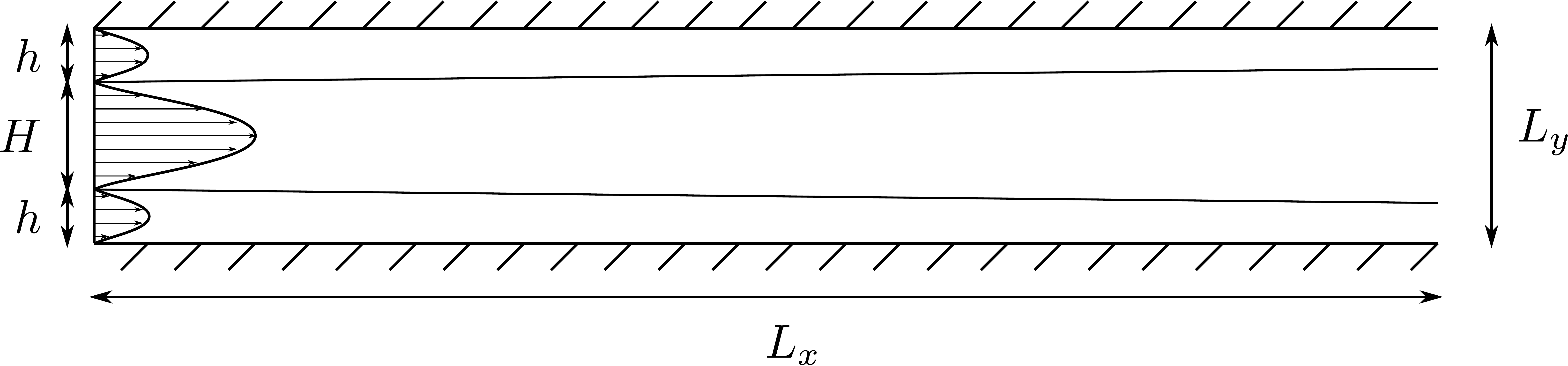}
   \caption{Illustration of the flow geometry.}
   \label{fig:geom}
\end{center}
\end{figure}

The flow considered is that of two immiscible fluids in a planar channel as depicted on Fig.\ \ref{fig:geom}. In this configuration, an inner fluid stream (fluid 1) is symmetrically sandwiched between two outer streams of fluid 2. The geometry and inflow are symmetric with respect to the centerline ($y=0$). In the following, $H$ denotes the channel height, $\bar{U}_1$ and $\bar{U}_2$ the bulk velocity of fluid 1 and 2 at the inflow, $\mu_1$ and $\mu_2$ their viscosities, and $\gamma$ the surface tension between them. 
 
Varying all relevant parameters would result in a huge number of simulations, and hence several of them are fixed. The half height of the inner fluid is fixed to $H/2$, which means that the confinement parameter (denoted by $h$ in~\cite{KTHwakes,StableWe} and not to be confused with the interfacial height $h$ in this work) is $1.0$ throughout this work. The fluids have the same uniform density $\rho$. The shear ratio $\Lambda^{-1} = (\bar{U}_1 + \bar{U}_2)/(\bar{U}_1 - \bar{U}_2)$, with $\bar{U}$ the average inflow velocity of each layer, will be fixed to $\Lambda^{-1}=1.4$.  The flow case considered is thus a co-flow jet~\cite{StableWe} - the inner flow stream has the highest velocity. The average inflow velocity of the outer fluid is 17$\%$ of that of the inner fluid. 

The remaining parameters are:
\begin{itemize}
  \item The Reynolds number: $Re = \rho \bar{U}_1 H / 2\mu_1$.
  \item The viscosity ratio: $m = \mu_2 / \mu_1$.
  \item The Weber number: $We = \rho \bar{U}_1^2 H / 2\gamma$.
\end{itemize}
Varying all three parameters simultaneously in the DNS would still be very expensive and make the visualization of neutral surfaces in such a large parameter space complicated. Hence, we vary the Reynolds number and viscosity ratio, while the Weber number is fixed to $We=10$ in the DNS except in Sec.\ \ref{sec:infl}, but will be varied in the Couette flow model problem. It has to be noted that this jet is globally stable for all shear ratios without surface tension when $m=1$, but becomes globally unstable at moderate Reynolds numbers when $We=10$~\cite{StableWe}. 

\section{Governing equations and numerical methods}
\noindent The dynamics of this flow is governed by the two-fluid incompressible Navier-Stokes equations with jump conditions at the interface between the two fluids. The previous global instability studies~\cite{StableWe,VakarWe} were performed with a sharp interface approach with two different domains for the two fluids, with coupling conditions at the interface. The present DNS study is performed by a diffuse-interface approach. During this work we have developed an in-house solver OILS, which is based on the open-source solver TPLS (Two-Phase Level Set)~\cite{TPLS-web-page} which was successfully validated against local instability studies in~\cite{ONaraigh_JFM_2014}. TPLS/OILS uses a level-set approach along with a continuous surface tension model (see~\cite{Sussman_SIAM_1998}) to model the interface separating the two phases. The viscosity discontinuity and surface tension force are both smoothed over a region of 1.5 grid cells, and are continuous functions around the interface. In such a formalism, the Navier-Stokes equations governing the dynamics of the two-phase flow are given by:
~
\begin{equation}
  \left\{
  \begin{aligned}
   & \frac{\partial \phi}{\partial t} + {\bf U} \cdot \nabla \phi         = 0 \\
   & \frac{\partial {\bf U}}{\partial t} + ({\bf U} \cdot \nabla) {\bf U} = \\
   & -\nabla P + \frac{1}{Re} \nabla \cdot \left[ \mu \left( \nabla {\bf U} + \nabla {\bf U}^T \right) \right] + \frac{1}{We} \kappa \delta_{\epsilon}(\phi) {\bf n}\\
   & \nabla \cdot {\bf U}                                                 = 0
  \end{aligned}
  \right.
  \label{eq: Navier-Stokes level-set}
\end{equation}

\noindent where ${\bf U}$ and $P$ are the velocity and pressure fields, respectively. The function $\phi({\bf x},t)$ is the level-set function, indicating which fluid occupies the point ${\bf x}$ at a given instant of time $t$ ($\phi < 0$ for fluid 1, and $\phi > 0$ for fluid 2). Consequently, the height of the interface $h({\bf x},t)$ separating the two fluids is given by the zero level-set contour: $\phi({\bf x},t) = 0$. The level-set function is used to determine the unit vector ${\bf n}$ normal to the interface and the local curvature $\kappa$:

\begin{equation}
  \begin{aligned}
    {\bf n} = \frac{\nabla \phi}{\| \nabla \phi \|} \\
    \kappa = -\nabla \cdot {\bf n}.
  \end{aligned}
\end{equation}

The viscosity jump is expressed as:

\begin{equation}
  \mu = m ( 1 - H_{\epsilon}(\phi)) + H_{\epsilon}(\phi)
  \notag
\end{equation}

\noindent where $H_{\epsilon}(\phi)$ is a regularized Heaviside function smoothed across a width $\epsilon = 1.5 \Delta x$. Similarly, the function $\delta_{\epsilon}(\phi)$ in equation~\eqref{eq: Navier-Stokes level-set} is a regularized Dirac function with a compact spatial support on the interval $\left[ -\epsilon, \epsilon \right]$.

\subsubsection*{Computational domain and boundary conditions}

The computational domain (shown in Fig. \ref{fig:geom}) has the dimensions $\left[ 0, L_x \right] \times \left[-L_y/2, L_y/2 \right]$. In this study, $L_y=4$ is given by the nondimensionalisation. In the $x$-direction, the length was chosen to be $L_x=250$. This length was found to be sufficient for surface tension-driven linear global modes in~\cite{StableWe}, which was also confirmed by our initial DNS.

Several boundary conditions are needed in order to close the system of equations~\eqref{eq: Navier-Stokes level-set}. For the velocity, no-slip boundary conditions are imposed on both upper and lower walls of the channel, while a standard outflow boundary condition ({\it i.e.} $\partial_x {\bf U} = p = 0$) is prescribed at the outlet. The inlet velocity profile results from three Poiseuille streams joining at the inflow such that :

\begin{equation}
  U(y) = \left\{
    \begin{aligned}
      & \frac{3}{2} \frac{(\Lambda^{-1} - 1)}{(\Lambda^{-1} + 1)} ( 1 - 4(y - 1.5)^2) \text{ for } y > 1          \\
      & \frac{3}{2}(1 - y^2) \text{ for } -1 \le y \le 1 \\
      & \frac{3}{2} \frac{(\Lambda^{-1} - 1)}{(\Lambda^{-1} + 1)} ( 1 - 4(y + 1.5)^2)\text{ for }  y < -1
    \end{aligned}
  \right.
  \notag
\end{equation}

\noindent where $\Lambda=1.4$ is the shear ratio defined previously. Regarding the level-set function $\phi$, a Neumann boundary condition is imposed at the outlet while at the inlet

\begin{equation}
    \phi(y) = \left\{
      \begin{aligned}
        & 1 + z \text{ for } z < 0\\
        & 1 - z \text{ for } z \ge 0
      \end{aligned}
      \right.
    \notag
\end{equation}

\begin{figure}
  \centering
  \includegraphics[width=0.45\columnwidth]{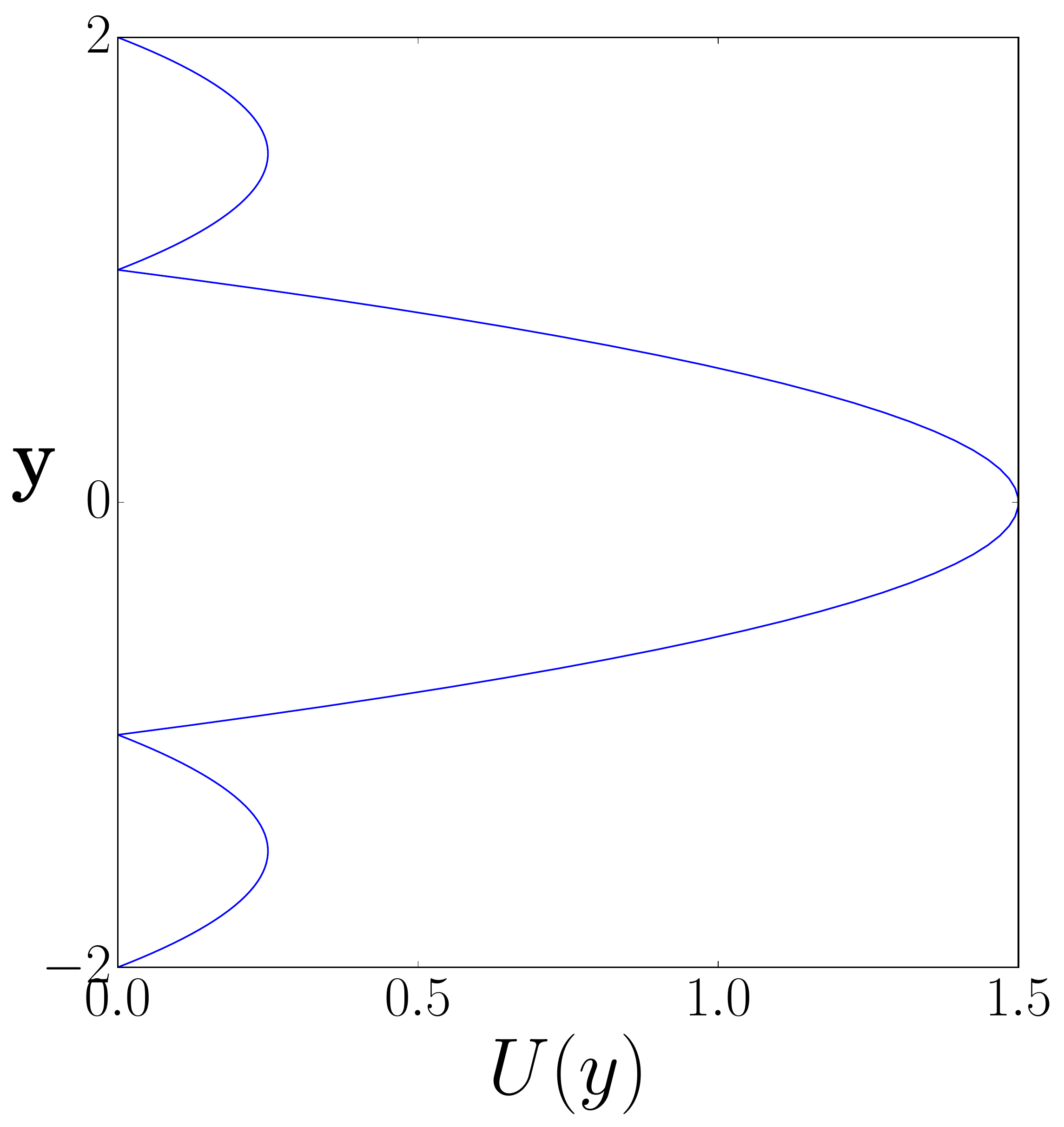}
  \includegraphics[width=0.45\columnwidth]{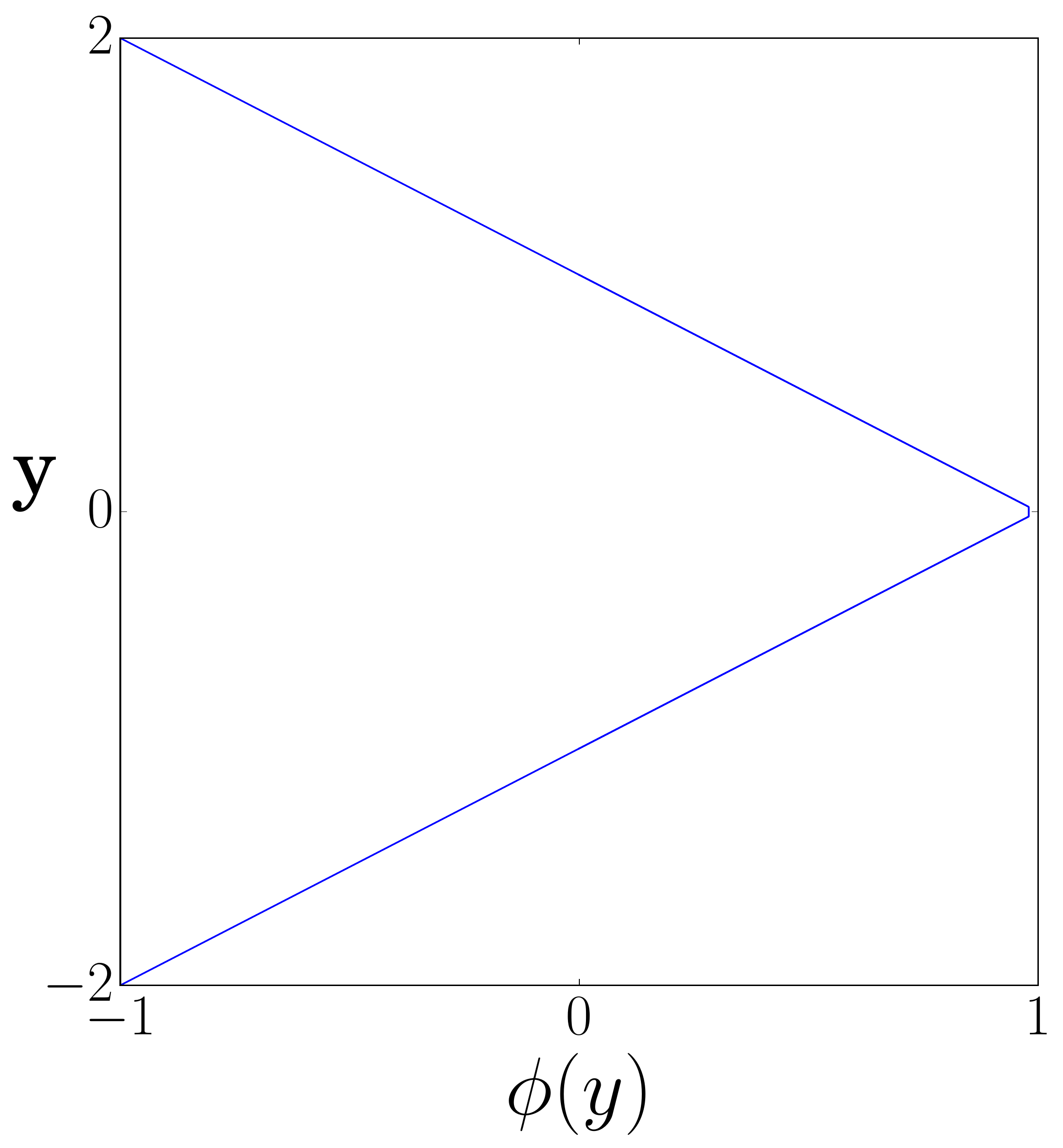}
  \caption{Illustration of the inflow profiles. (a) Inflow velocity profile for $\Lambda^{-1} = 1.4$ (b) Inflow level-set profile.}
  \label{fig:inflow_profiles}
\end{figure}

\noindent For the sake of illustration, Fig. \ref{fig:inflow_profiles} depicts the inlet velocity profile for $\Lambda^{-1}= 1.4$ along with the inlet level-set profile.
\subsubsection*{Discretization scheme}
The solver OILS uses the same finite difference discretization as TPLS~\cite{ONaraigh_JFM_2014}. The Navier-Stokes equations are discretized using a finite volumes method on a MAC grid with uniform grid spacing in all directions of space. The velocities are defined on the cell faces, while the scalars (level-set function $\phi$, pressure $P$, viscosity $\mu$) are defined at the cell centers. A fully explicit second order Adam-Basforth scheme is used for the temporal discretization of the Navier-Stokes equation and a Strong-Stability-Preserving Runge-Kutta 3 scheme (SSP-RK3) for the discretization of the level-set advection equation. The pressure and associated divergence-free constraint are treated using the projection method. A Poisson solver based on the Scheduled Relaxation Jacobi method~\cite{Xiyang_JCP_2014} has been used for the two-dimensional simulations in the present work, while the latest version of OILS instead contains a conjugated gradient solver preconditioned by the algebraic multigrid method. Finally, the level-set function $\phi({\bf x},t)$ is the signed-distance function such that $\| \nabla \phi \| = 1$ and is advected using the HOUC5 scheme~\cite{HOUC5}. The re-distancing of the the resulting level-set function is performed using the PDE-based approach and the algorithm of Sussman \& Fatemi~\cite{Sussman_SIAM_1998}. As for the advection of the level-set field, the pseudo-time discretization is based on the SSP-RK3 scheme while the spatial discretization now relies on a WENO5 scheme.

Regarding resolution, in initial studies, 128 points in the wall-normal direction (resulting in a grid spacing of $\delta x= \delta y=7.8 \cdot 10^{-3}$) were found to be sufficient for most parameter values, and used throughout this work, except for the lowest viscosity ratio ($m=0.1$), for which 256 points in the wall-normal direction ($\delta x=\delta y=3.9 \cdot 10^{-3}$) were needed to fully capture the details of the interfacial waves and the wall boundary layer. 

\section{Results}
\subsection{Presence of a global instability in DNS for uniform viscosity co-flow jets ($m=1$)}

The first study to be performed is to confirm that the surface tension-induced instability of co-flow jets~\cite{StableWe} found by linear global mode analysis also appears in nonlinear simulations (DNS). If very close to a neutral stability boundary, the flow in the DNS first turns towards the \textit{base flow}, which is a steady solution to the Navier-Stokes equations (here expressed in the level-set formalism):
~
\begin{equation}
  \left\{
  \begin{aligned}
   & {\bf U}_{b} \cdot \nabla \phi_b         = 0 \\
   &  ({\bf U}_b \cdot \nabla) {\bf U}_b = \\
   & -\nabla P_b + \frac{1}{Re} \nabla \cdot \left[ \mu \left( \nabla {\bf U}_b + \nabla {\bf U}_b^T \right) \right] + \frac{1}{We} \kappa \delta_{\epsilon}(\phi)_b {\bf n}\\
   & \nabla \cdot {\bf U}_b                                                 = 0
  \end{aligned}
  \right.
  \label{eq: Navier-Stokes level-set steady}
\end{equation}

Close to the neutral global stability limit, and if convective instabilities are not too strong, the appearance of global instability can be quantified in DNS by looking at time traces of the velocity. When an unstable global mode is present, the DNS time signal in any given point in space grows exponentially in time. If the global modes are all stable ($\sigma_r<0$ for all modes), then the time trace should exhibit an exponential decay.

\begin{figure}
  \centering
  \includegraphics[width=0.8\columnwidth]{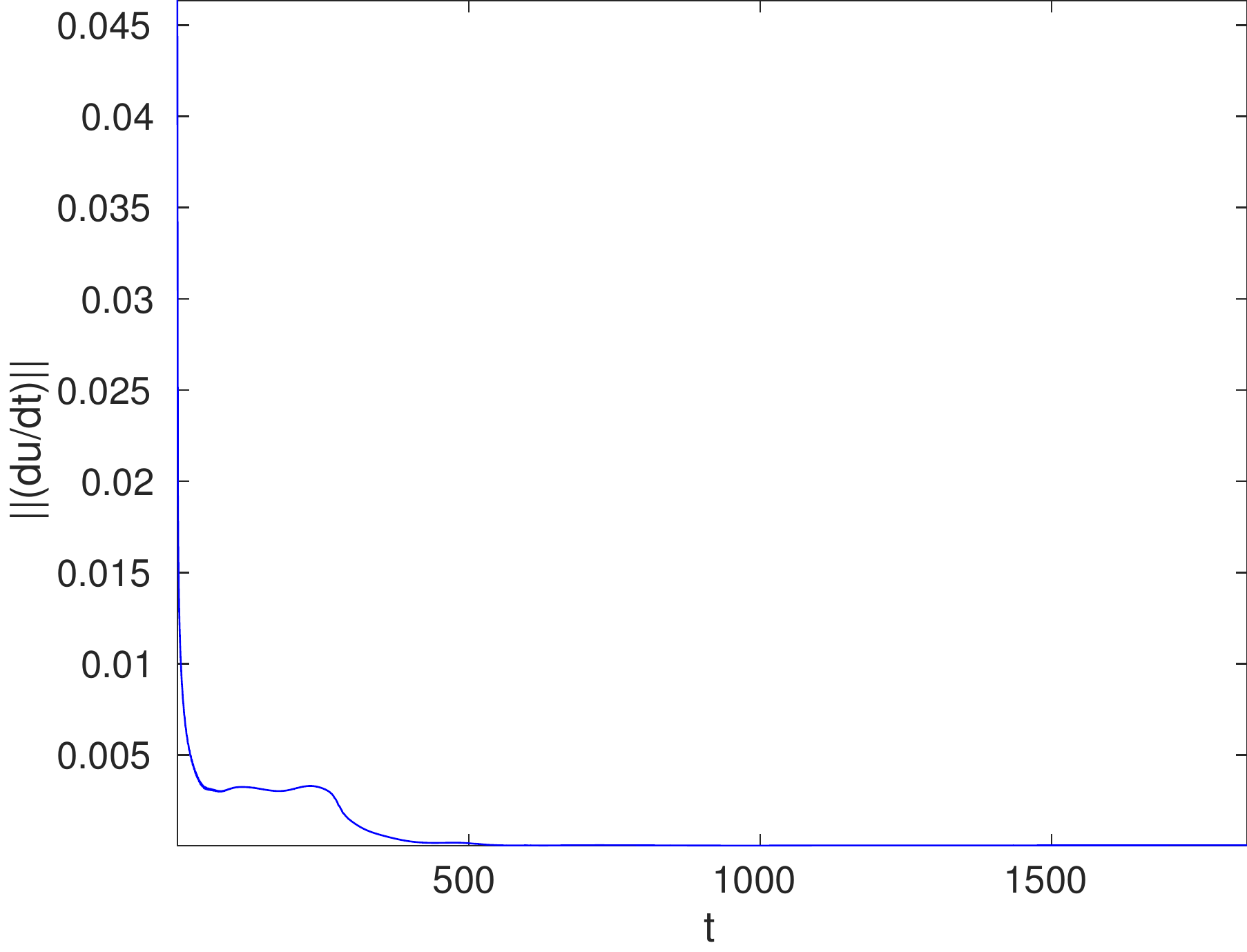}
   \includegraphics[width=0.8\columnwidth]{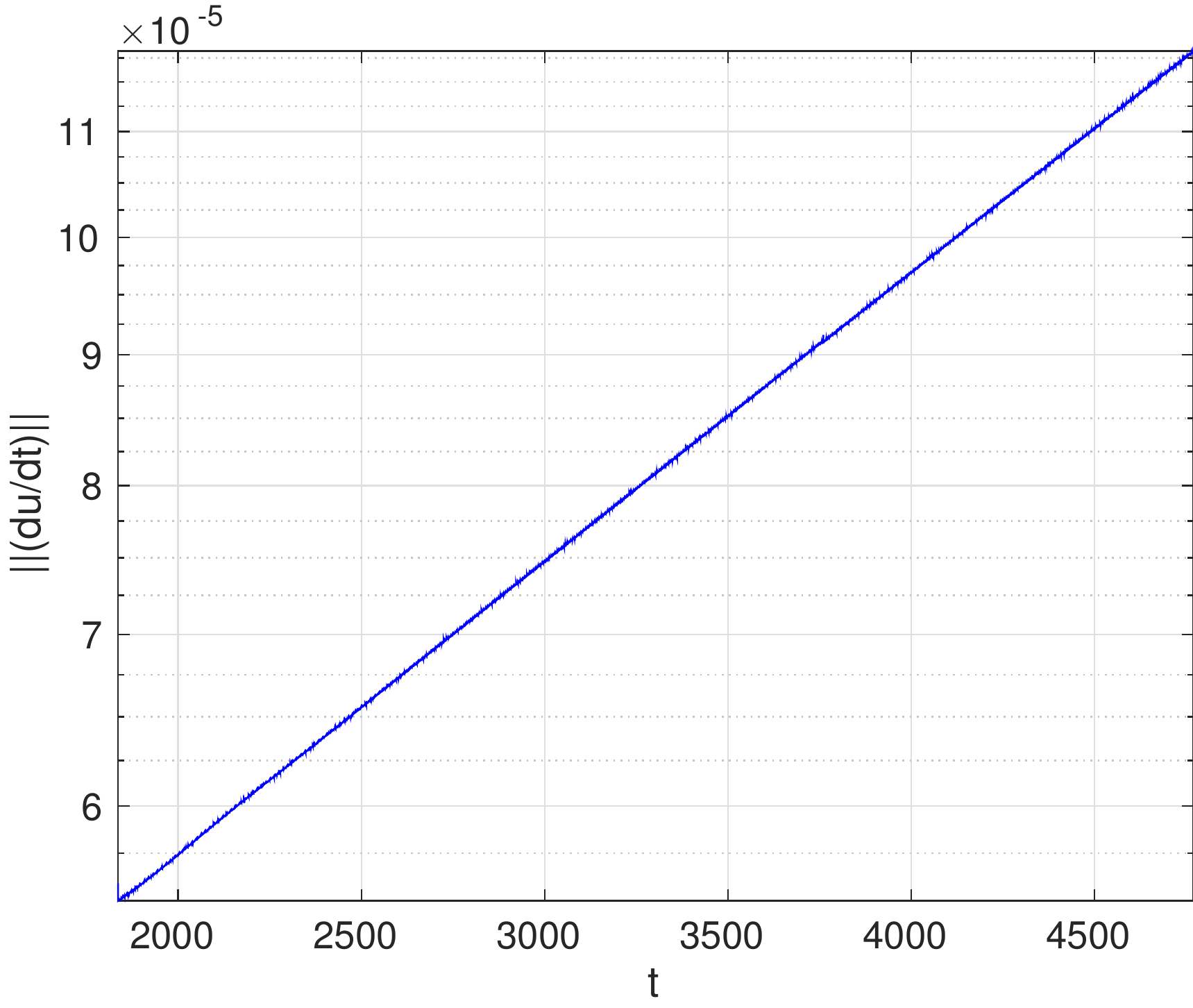}
  \caption{Top: Norm of the average time derivative over the flow domain over time, $m=1$, $Re=250$. This shows the initial decay towards steady state. Bottom: The same but at later times and in logarithmic ($y$-)scale. This shows the initial exponential growth of a linear global mode, which will saturate to a constant-amplitude limit cycle at later times.}
  \label{fig:m1Re250resplot1}
\end{figure}

In this study, time-dependent oscillations are quantified by recording the spatial average of the time derivative of the velocity magnitude over the whole computational domain, simply termed the \textit{residual} in the rest of this study, given by:
\begin{equation}
  {(L_x L_y)^{-1}\int_D ||\textbf{u}(t+\delta t)-\textbf{u}(t))||/\delta t) }
\end{equation}

\noindent where $\int_D$ denotes an integration over the whole computational domain. This residual is shown for $m=1$, $Re=250$ in Fig.\ref{fig:m1Re250resplot1}. It shows a clear initial decay towards a steady state $U_b$ (top), and a later exponential growth in the vicinity of the steady state (bottom). Such observation indicates that a linear global mode is growing in the DNS, and the flow at $Re=250$ is hence globally unstable. A similar study for $Re=245$ shows an exponentially decaying residual, indicating that the flow in DNS is stable. A linear interpolation between the two growth/decay rates gives a neutral point and the critical Reynolds number $Re_c=248$. 

\begin{figure}
  \centering
  \includegraphics[width=\columnwidth]{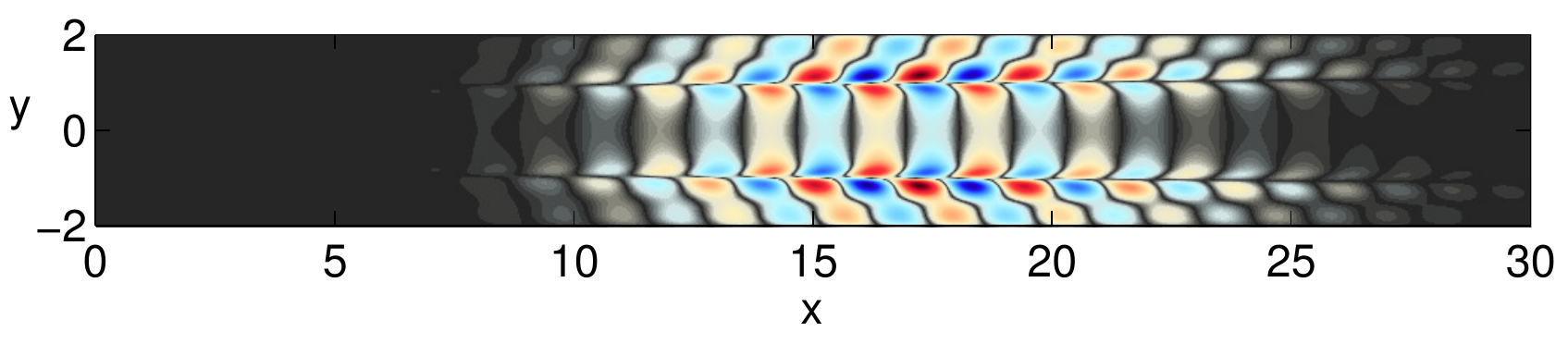}
  \includegraphics[width=\columnwidth]{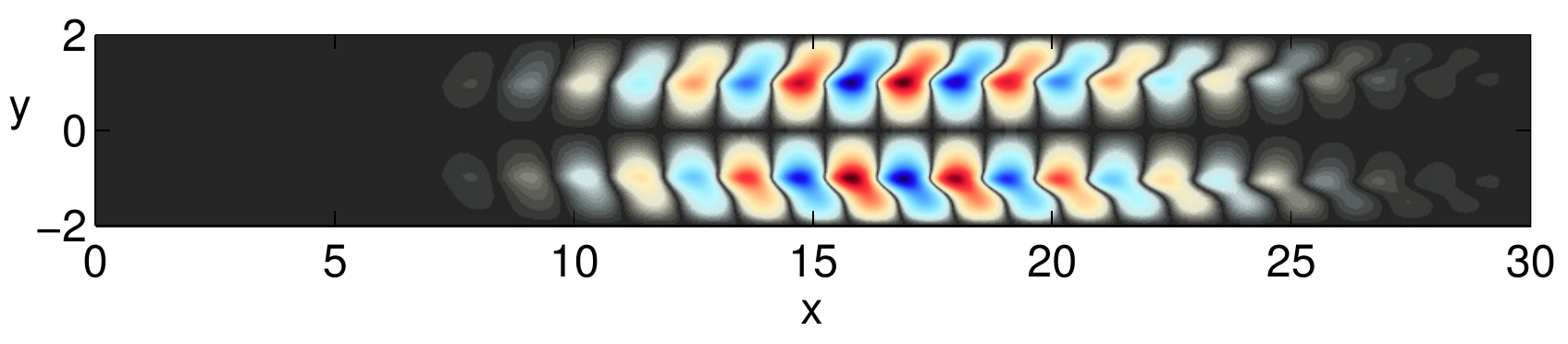}
  \caption{The difference between the instantaneous velocity from DNS and the steady solution, at $m=1$, $Re=250$, during the exponential growth phase. Top: streamwise velocity field, Bottom: vertical velocity field.}
  \label{fig:Wmode_m1_growing}
\end{figure}

\begin{figure}
  \centering
  \includegraphics[width=\columnwidth]{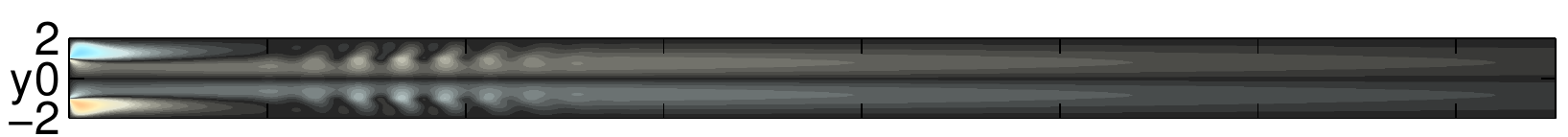}
  \includegraphics[width=\columnwidth]{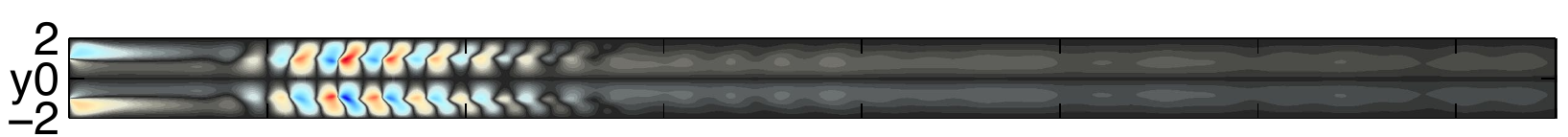}
  \includegraphics[width=\columnwidth]{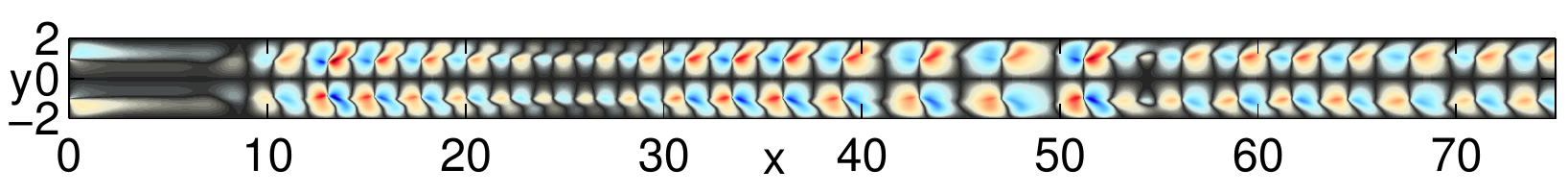}
   \caption{The vertical velocity field at the saturated nonlinear state, $m=1$. Top: $Re=250$, middle: $Re=316$, bottom: $Re=500$. The colorscale limits are $\pm0.1$ in all figures.}
  \label{fig:Wmode_m1_long}
\end{figure}

\begin{figure}
  \centering
  \includegraphics[width=\columnwidth]{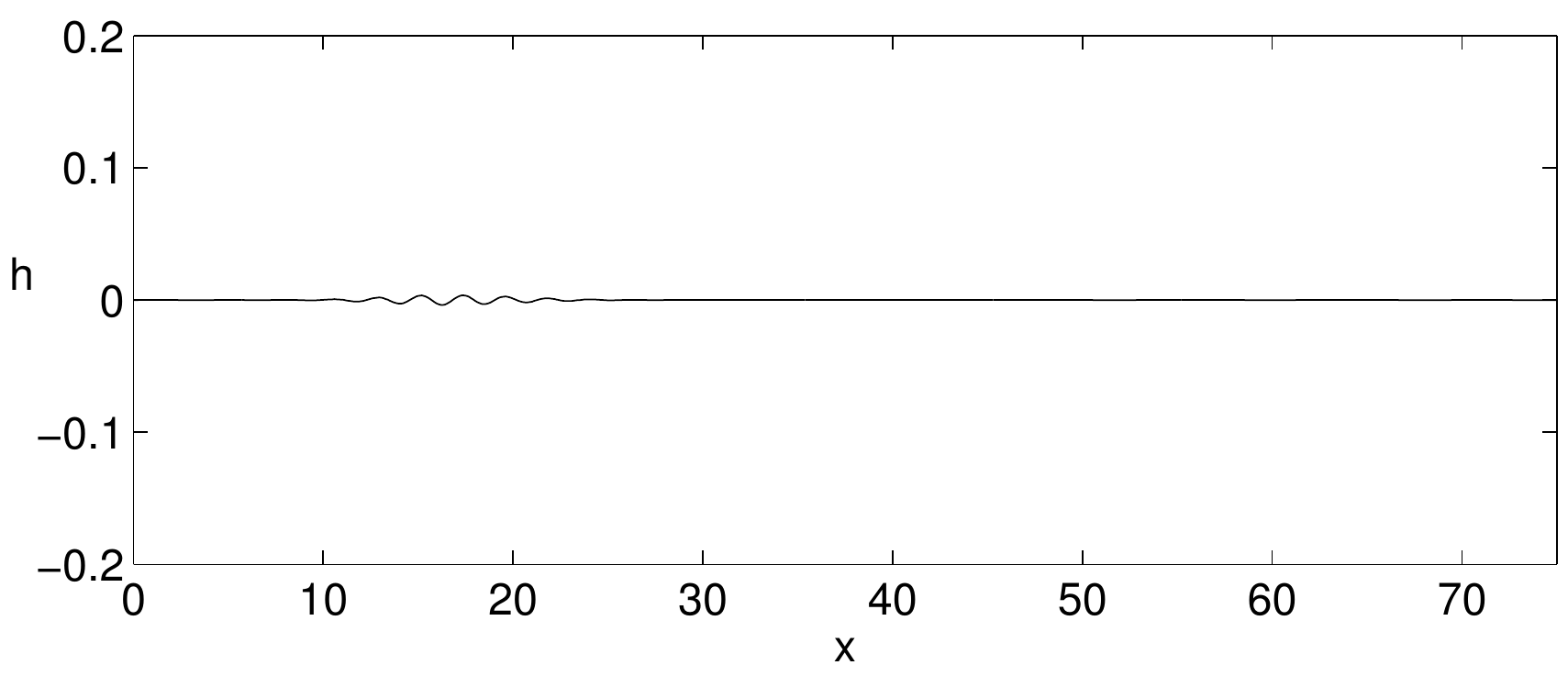}
  \includegraphics[width=\columnwidth]{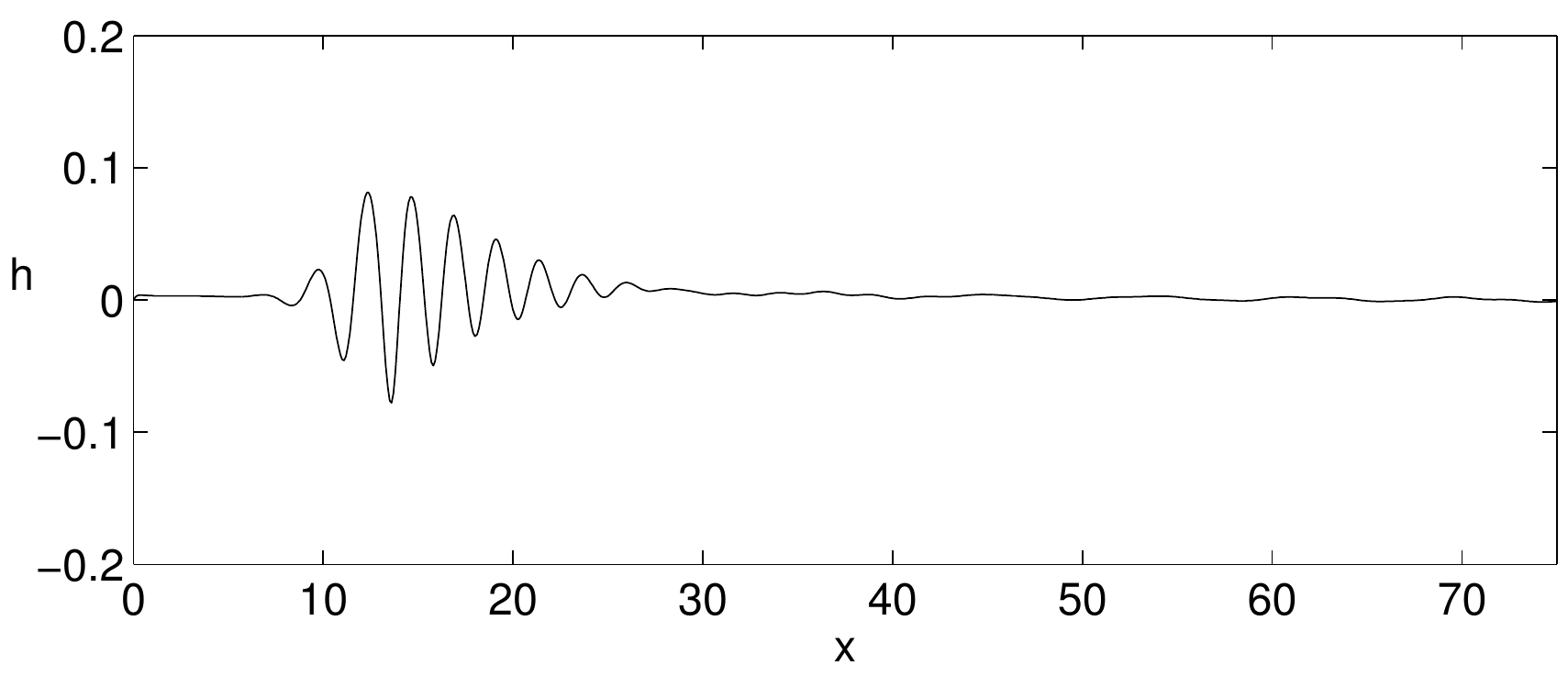}
  \includegraphics[width=\columnwidth]{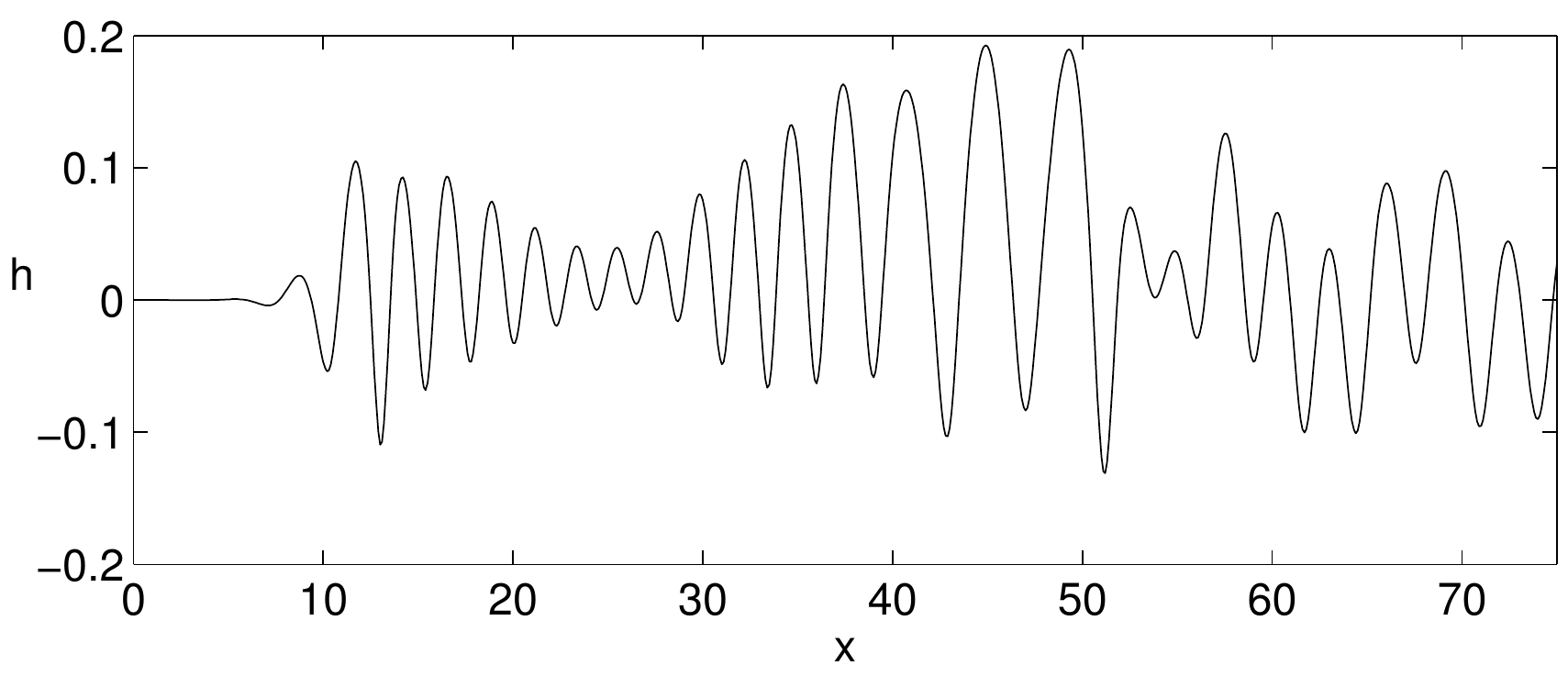}
   \caption{The instantaneous (upper) interface position, mean value subtracted, at the saturated nonlinear state, $m=1$. Top: $Re=250$, middle: $Re=316$, bottom: $Re=500$.}
  \label{fig:hmode_m1_long}
\end{figure}

The growing eigenmode is depicted in Fig.\ \ref{fig:Wmode_m1_growing}. The oscillation displays short-wavelength waves localized around the interface, in the upstream part of the computational domain. The unstable global eigenmodes for the jet in figure 6 of~\cite{StableWe} (at $\Lambda^{-1}=1.2$ and $Re=316$) also had short-wavelength waves around the interface at a similar wavelength. Knowing that both jets are globally stable without surface tension (observed in the present work as well as in~\cite{StableWe}), the resemblance strongly indicates that we are observing the same surface tension-induced global instability. It is also worth investigating whether the jet modes saturate at a very low level and without visible interface perturbation, as was indicated especially for the varicose wake mode in~\cite{Biancofiore_FDR_2014}. First, figure \ref{fig:Wmode_m1_long} shows the vertical velocity ($V(x,y)$) of the final oscillatory state at $m=1$ (without subtracting the base flow), at increasing Reynolds numbers, from $Re=250$ at the top to $Re=500$ at the bottom. When Reynolds number increases, the amplitude of the vertical velocity oscillation increases significantly, and the mode becomes much more elongated in the streamwise direction. The interfacial perturbation amplitudes at different Reynolds numbers can be compared in figure \ref{fig:hmode_m1_long}. At bifurcation ($Re=250$, top), the maximal interface displacement is $|h|=0.004$, which is barely visible, and comparable to the wake modes in~\cite{Biancofiore_FDR_2014} at $|h|=0.01$. At $Re=500$, the global instability perturbs the interface significantly - up to $|h|=0.19$. 


\subsection{Local stability analysis}

It is instructive to find out where the absolute instability driving the global mode is located. A local spatiotemporal analysis has been performed for the jet flow slightly above the onset of instability: $m=1$, $Re=250$. Exactly one absolutely unstable mode was found; this is a hidden neutral mode destabilized by surface tension, called the interfacial mode in~\cite{VakarWe,StableWe}. The mode is symmetric (varicose), in agreement with the DNS shown in the previous subsection. 

\begin{figure}
  \centering
  \includegraphics[width=\columnwidth]{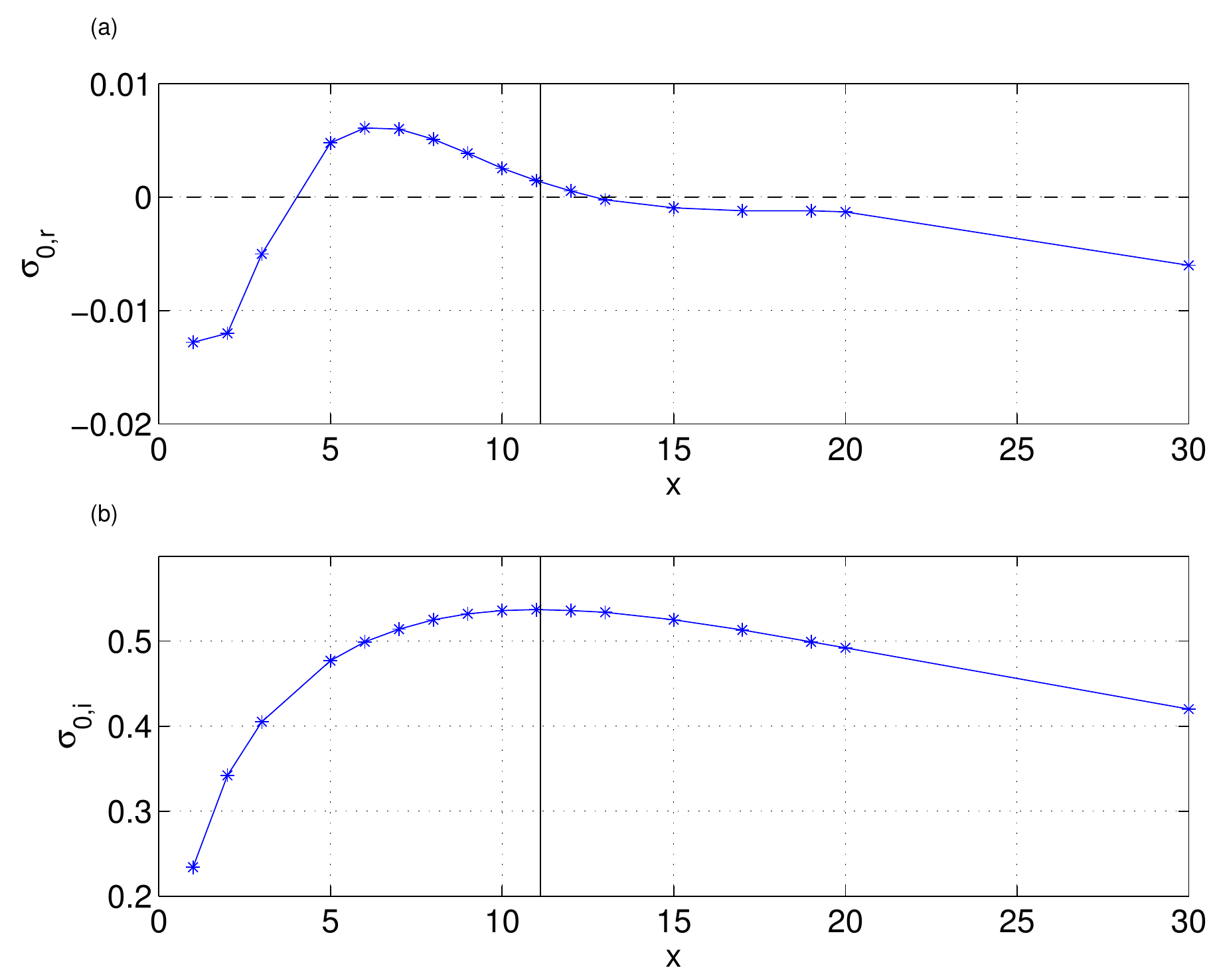}
  \caption{The local absolute frequency of the base flow computed by TPLS at $m=1$, $Re=250$: (a) absolute growth rate $\sigma_{0,r}$, (b) absolute frequency $\sigma_{0,i}$. The vertical line indicates the position of the saddle in the complex $X$-plane.}
  \label{fig:m1Re250_omega0}
\end{figure}

The streamwise evolution of the local absolute growth rate is shown in figure \ref{fig:m1Re250_omega0} (a), while that of the local absolute frequency is presented on figure \ref{fig:m1Re250_omega0} (b). The flow displays a pocket of absolute instability between $x=4$ and $x=13$. The maximum absolute growth $\sigma_{r,max}=0.06$ occurs at $x=6$. The approximate global mode frequency and wavemaker position can be found by an analytic continuation of $\sigma_0$ to the complex $X$-plane. This is done here by fitting Pad$\acute{\textrm{e}}$ polynomials around the point of maximum absolute growth, as in~\cite{juniperwakes}. The linear global mode frequency approximated by local spatiotemporal analysis this way is $\sigma_{g,l}=0.0013+0.54i$. The angular frequency extracted from the DNS time signal during the exponential growth is $0.53$, in very good agreement with the local analysis. 

\begin{figure}
  \centering
  \includegraphics[width=\columnwidth]{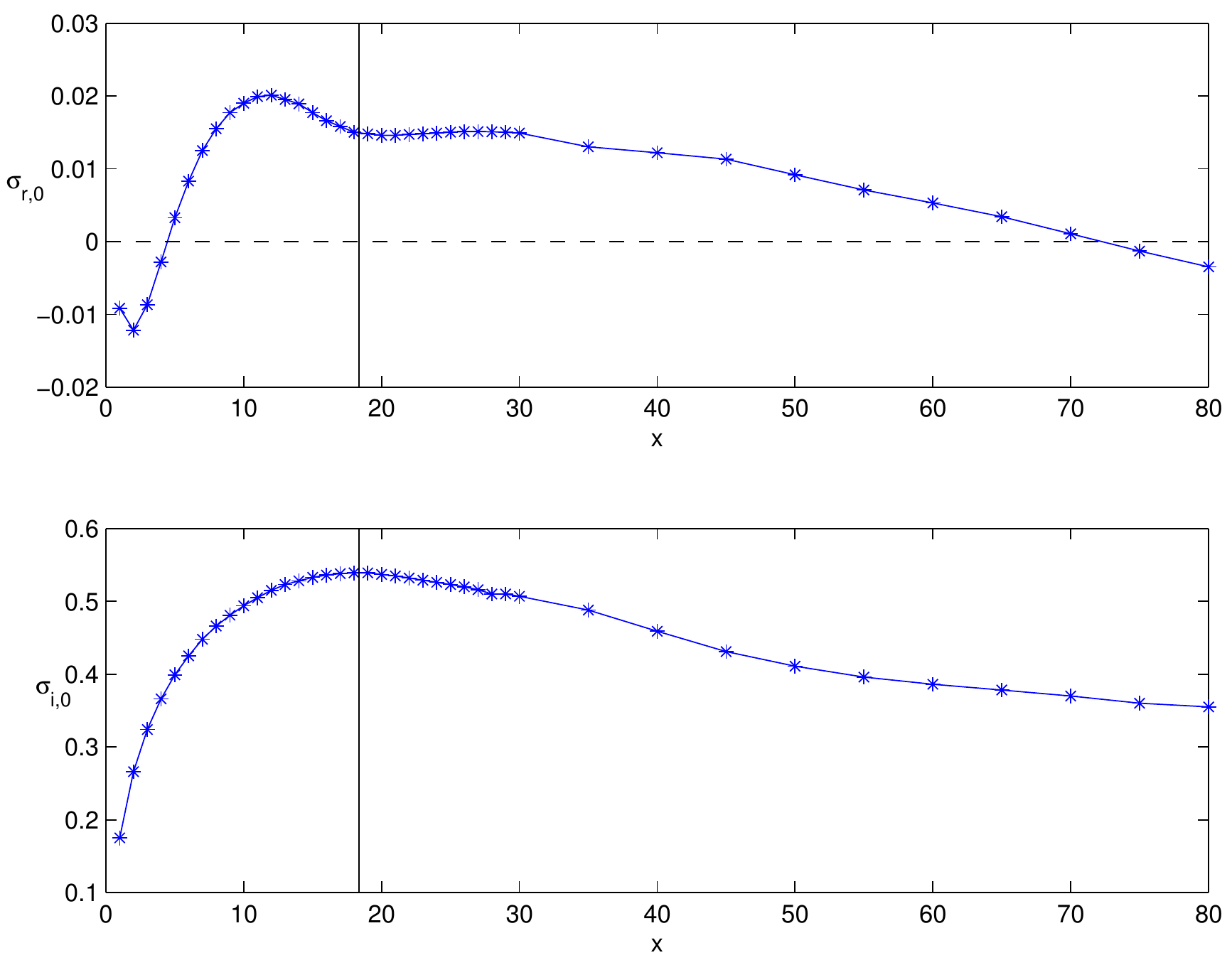}
  \caption{The same as in figure \ref{fig:m1Re250_omega0} but at $m=1$, $Re=500$.}
  \label{fig:m1Re500_omega0}
\end{figure}

The nonlinear oscillation waves observed in figure \ref{fig:Wmode_m1_long} all have a similar envelope upstream in the domain. Further downstream, the modes at $Re=250$ and $Re=316$ (figure \ref{fig:Wmode_m1_long}, top) decay and have negligible amplitudes for $x>30$. The mode at $Re=500$ (Fig. 6 bottom) on the other hand grows again at $x\approx 30$ and remains at a large amplitude at $x=75$. In~\cite{StableWe}, similar very elongated jet modes were obtained for a range of parameter values for which a coupling was occurring between the upstream absolute instability pocket and a convective instability pocket downstream (the convective instability having "accidentally" the same frequency as the absolute instability). It therefore deserves to be investigated whether the second growth region is due to an absolute or convective instability\footnote{It should be noted that one should expect only indicative relation between the nonlinear oscillation shape and absolute instability regions this far from bifurcation}. Figure \ref{fig:m1Re500_omega0} shows the absolute instability at $Re=500$. This reveals a similar upstream onset of absolute instability as at $Re=250$, at the same frequency 
($\sigma_{g,l}=0.016+0.54\textrm{i}$), but the absolute growth at $Re=500$ does not decay. The flow instead exhibits an extremely long pocket of absolute instability - until $x\approx75$. Hence, the long mode observed for $Re=500$ is due to a persistent absolute instability, in contrast to the elongated modes in~\cite{StableWe} which arose through a coupling of an absolute instability pocket with a second convective instability mode. Long modes of high amplitude can thus arise due to several different mechanisms, further highlighting the rich dynamics exhibited by such a basic flow case as immiscible planar jets with surface tension.

\begin{figure}
  \centering
  \includegraphics[width=\columnwidth]{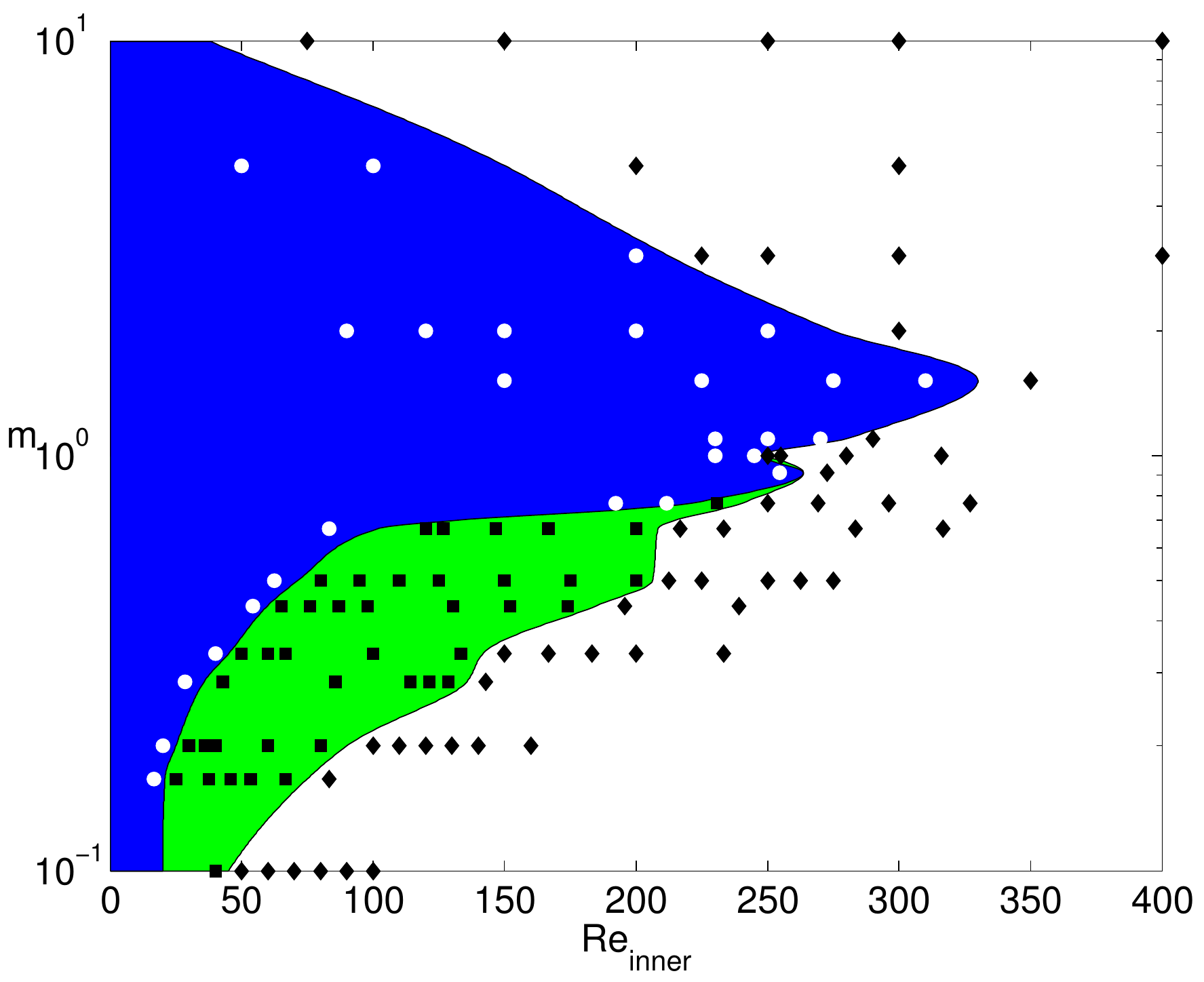}
  \caption{Neutral stability limit for persistent instability in DNS as a function of the inner flow Reynolds number. In the blue region the flow is stable, in the green region the flow displays a steady Coanda attachment, the white region the flow displays time-dependent oscillations. The markers show all the DNS runs, with black marker (on white) denoting persistent oscillation and white marker (on grey) decay of the oscillations (residual settled at a level below $10^{-5}$.)}
  \label{fig:Re_inner}
\end{figure}
\begin{figure}
  \centering
  \includegraphics[width=\columnwidth]{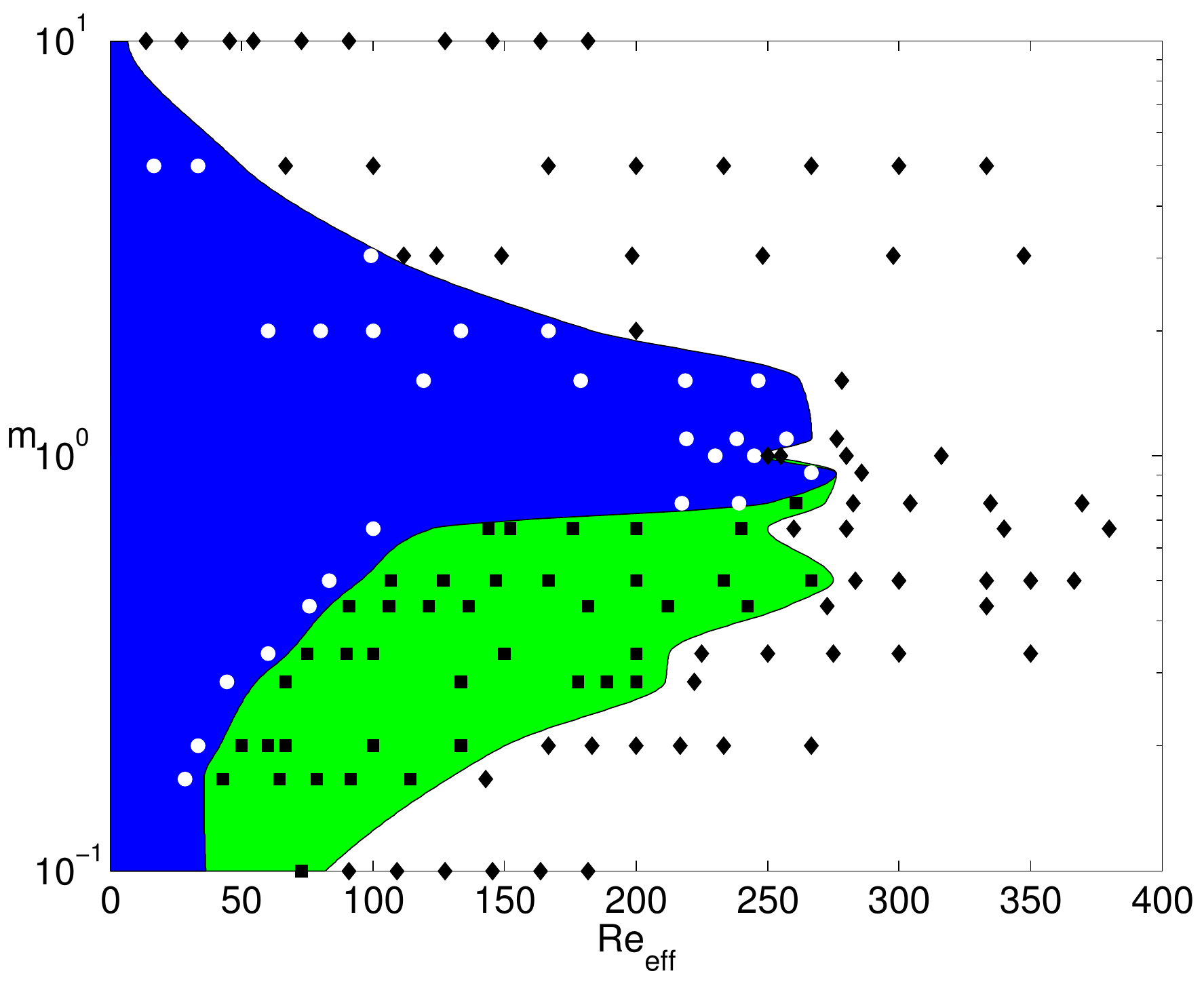}
  \caption{The same neutral curves and data as in Fig.\ \ref{fig:Re_inner}, but shown as a function of the Reynolds number based on average viscosity.}
  \label{fig:Re_eff}
\end{figure}

%
\begin{figure}
  \centering
  \includegraphics[width=\columnwidth]{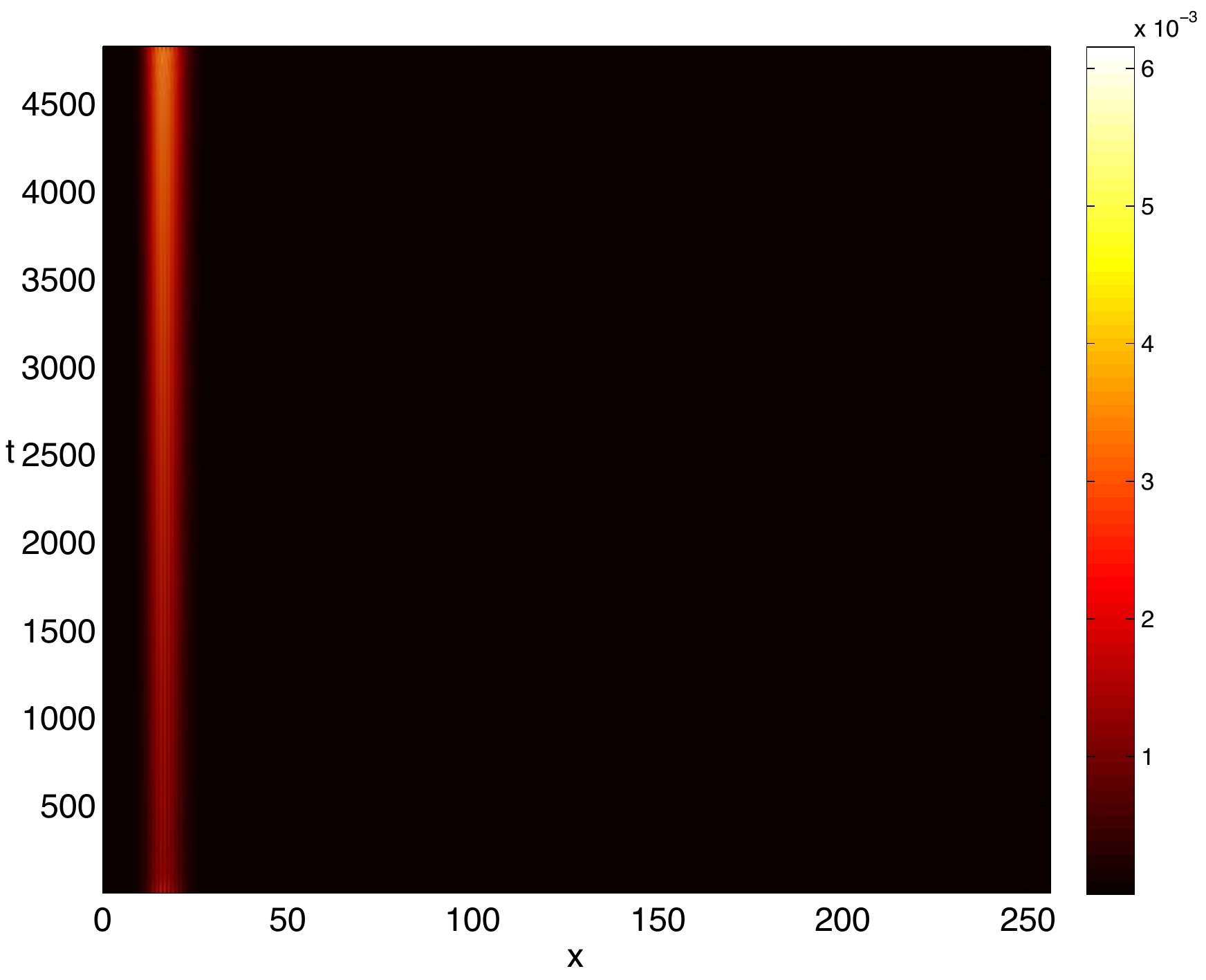}
  \caption{A space-time diagram of the kinetic energy of the antisymmetric perturbation along the lines $y=\pm1$, at $m=1$, $Re=250$. The perturbation grows at the location of the source, which is a sign of global (~absolute) instability. }
  \label{fig:m1Re250_spacetime}
\end{figure}

\subsection{Effect of the viscosity ratio on the instability}

The influence of the viscosity ratio on the instability and transition to unsteadiness is now investigated. We note that for non-uniform viscosity, we could not always observe a clear exponential growth in our time signals. Strong convective instability bursts could be masking a slow exponential growth in time, especially far away from instability boundaries (unknown a priori). Hence, our classification of stable/steady and oscillatory flow cases is based on the final saturated flow state. To obtain the neutral curve in the $Re$-$m$-plane, we have applied the following procedure. First, at each viscosity ratio, we have scanned a range of Reynolds numbers in DNS to approximately locate the neutral curve. Closer to the neutral curve, a steady solution for Navier--Stokes equation (base flow) is obtained using selective frequency damping~\cite{sfd}. Then, a new DNS is started from the base flow and the residual and perturbation norm recorded over a period of at least $1000$ (but typically $>3000$) non-dimensional time units. 
At this point, transients have usually decayed and the state of the flow is "statistically steady" - \textit{i.e.} the average residual over $100$ consecutive time units is constant. It should be mentioned that the residual is a time-derivative, and hence may be higher than the actual perturbation amplitude if high-frequency numerical noise is present. In this work, the following threshold has been adopted: if the final residual and the perturbation amplitude both have settled at a level larger than $10^{-5}$,  then the flow is classified as unsteady, otherwise, the flow is classified as steady. A table of all simulation times, parameters, and final residual levels is included in Appendix \ref{app:tab}.

Fifteen different values of the viscosity ratio have been considered, ranging from $m=0.1$ (outer fluid much less viscous than inner fluid) to $m=10$ (outer fluid much more viscous than the inner fluid). Regions of steady and oscillatory solutions in the $Re$-$m$-plane are shown in figure \ref{fig:Re_inner}, as a function of the inner flow Reynolds number. Inside the blue region, the final flow state is stable (steady and symmetric). Inside the white region, the final flow state is unsteady. Inside the green region, the final flow state is steady but asymmetric. The different sets of parameters considered are all depicted by markers. 

Figure \ref{fig:Re_inner} shows that a small viscosity contrast in any direction (the outer fluid more viscous or less viscous) is stabilizing. This indicates that the surface tension-induced global instability is stabilized by a viscosity contrast in any direction. However, the figure clearly highlights that critical Reynolds number decreases with viscosity ratio. This indicates that other instability mechanisms are active. 
The highest critical Reynolds number (the most stable case) $Re_c\approx330$, is achieved for a viscosity ratio $m=1.5$, \textit{i.e.} when the outer fluid is slightly more viscous than the inner. 

Finally, the same trends are observed when the Reynolds number is based on the average viscosity $(\mu_1+\mu_2)/2$), shown in figure \ref{fig:Re_eff}. It is worth noting moreover that the effective Reynolds number at the onset of instability is not constant for different $m$. Hence, not only does the viscosity ratio changes the effective critical Reynolds number, but it also strongly influences the dominant instability mechanisms. 

In the following, the instability mechanisms with more viscous outer fluid ($m>1$) and less viscous outer fluid ($m<1$) are analysed separately. To examine the nature of the instabilities - absolute or convective - we instead rely on space-time diagrams, similarly to~\cite{Biancofiore:2010p1006} who used such figures to find global instabilities for confined wakes in DNS.  The space-time diagram for the supercritical global instability at $m=1$, $Re=250$ is shown in figure \ref{fig:m1Re250_spacetime}. The quantity depicted is the sum of the local kinetic energies along the lines $y=1$ and $y=-1$. The region of the global mode (and absolute instability) is distinctively picked out as a vertical line. The lines of constant amplitude are all vertical, showing that when time increases, the amplitude stays constant in space.

\begin{figure}
  \centering
  \includegraphics[width=\columnwidth]{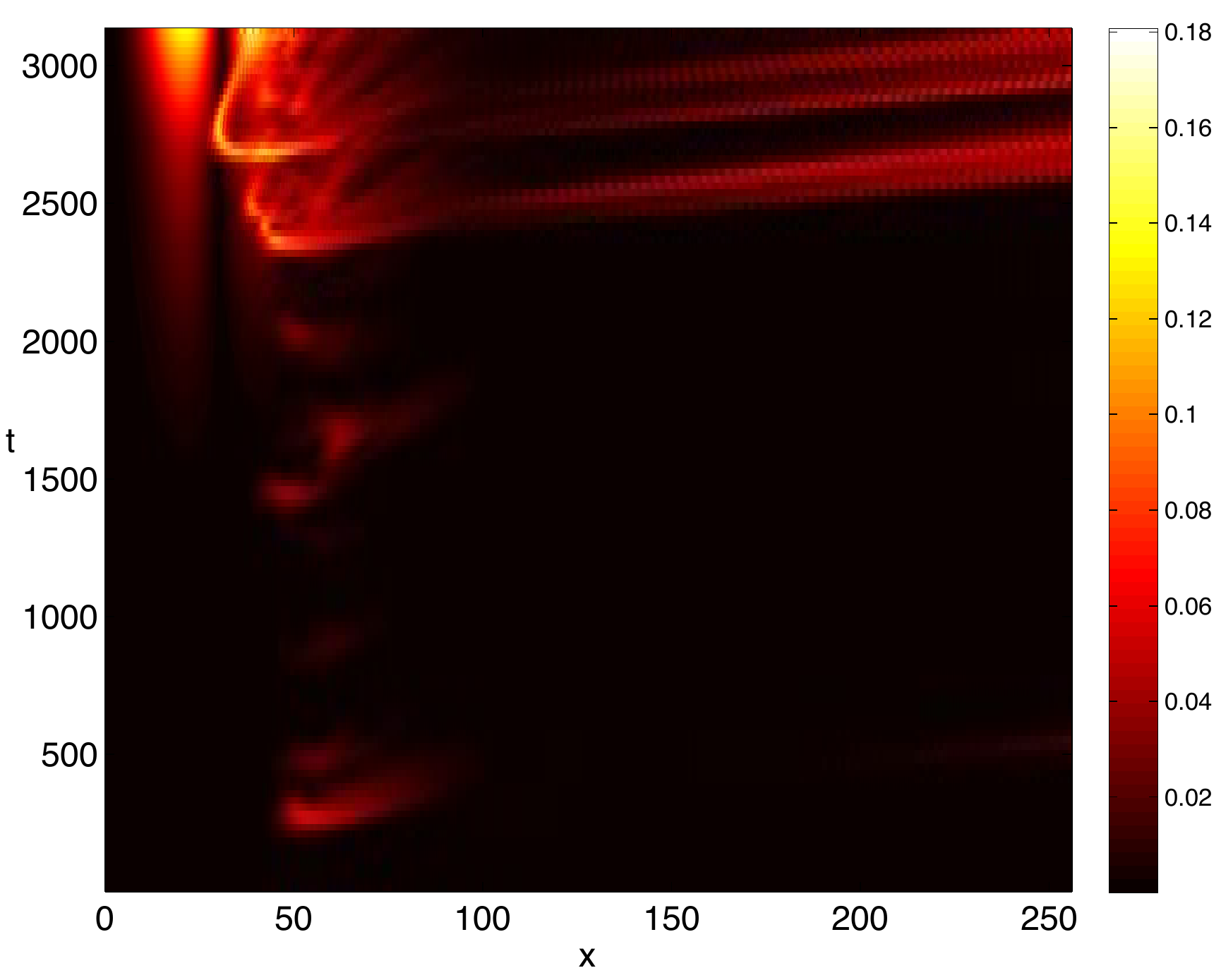}
  \caption{A space-time diagram of the kinetic energy of the antisymmetric perturbation along the lines $y=\pm1$, at $m=0.5$, $Re=250$. This reveals an upstream region of global instability followed by (and triggering) convective bursts downstream.}
  \label{fig:m2Re500_spacetime}
\end{figure}

\begin{figure}
  \centering
  \includegraphics[width=\columnwidth]{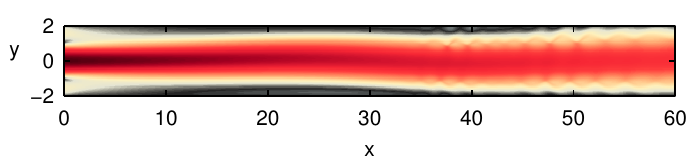}
  \includegraphics[width=\columnwidth]{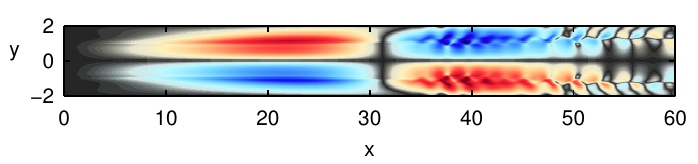}
  \caption{The streamwise velocity of the flow at $m=0.5$, $Re=250$ at $t=3100$ (the uppermost part of the space-time diagram in figure \ref{fig:m2Re500_spacetime}.). Top: the horizontal velocity field, bottom: the antisymmetric part of the same. Note that the $y$-axis is magnified by a factor $2.5$ compared to the $x$-axis.}
  \label{fig:m2Re500_antisymmetric}
\end{figure}

\begin{figure}
  \centering
   \includegraphics[width=\columnwidth]{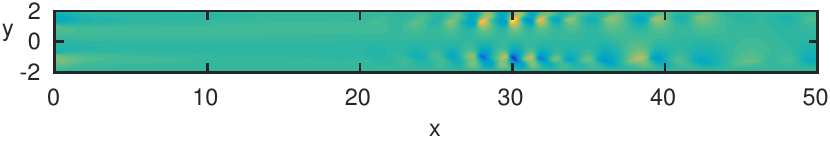}
   \includegraphics[width=\columnwidth]{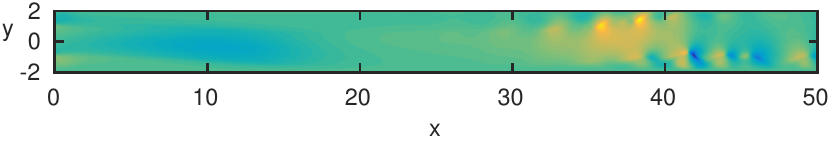}
  \includegraphics[width=\columnwidth]{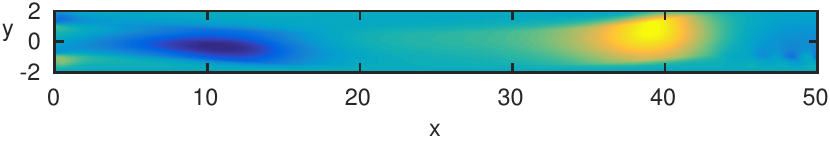}
  \caption{The instantaneous vertical velocity DNS at $m=5$, $Re=500$, time increasing from top to bottom. The jet shows initially short-wavelength waves similar to the uniform density jet, but finally arrives at a steady Coanda attachment.}
  \label{fig:m0dot2Re500_withtime}
  \end{figure}

\begin{figure}
  \centering  
  \includegraphics[width=\columnwidth]{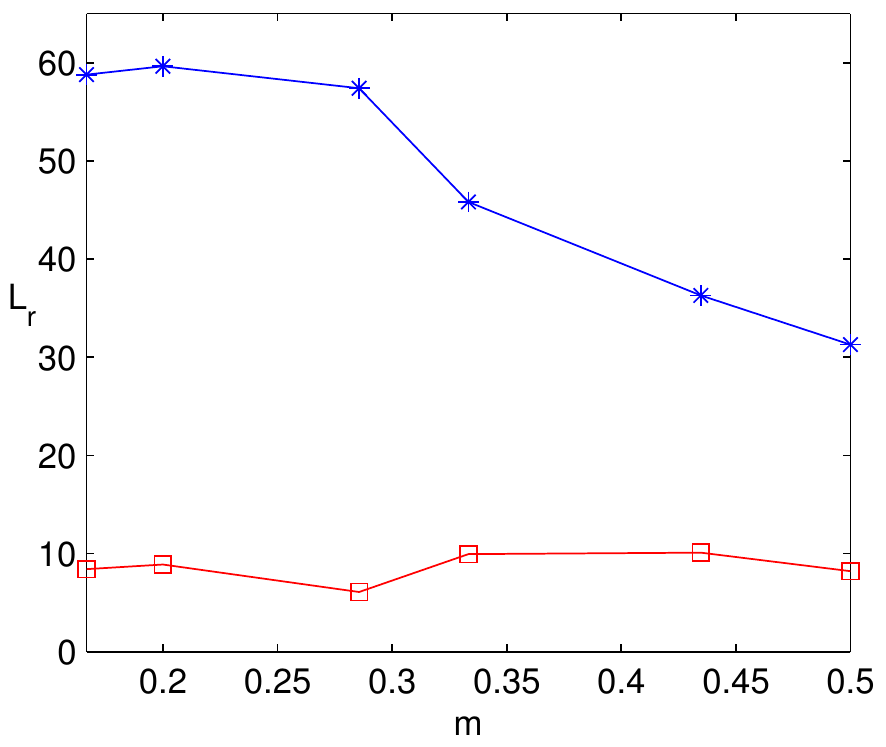}
  \includegraphics[width=\columnwidth]{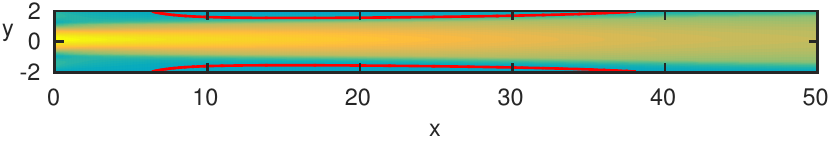}
  \caption{Top: Recirculation length at constant $Re=230$ as a function of $m$ (line with stars, blue online), and at the onset of Coanda attachment (line with squares, red online). Bottom: Streamwise velocity at an early time (before the onset of oscillations or asymmetry). Zero streamwise velocity contour shown in red to emphasize the extent of the recirculation zones.}
  \label{fig:Recircvsm}
\end{figure}

\subsection{Low viscosity of the outer fluid}
At $m=0.9-1.1$, when the Reynolds number is increased from zero, the first bifurcation is through the surface tension-induced global mode at $Re\approx 250$. At low enough viscosity of the outer fluid however (approx. $m<0.9$), a different scenario emerges which is described below.

A space-time diagram for an oscillatory state at $Re=250$ (inside the white region) at $m=0.5$, is shown in figure \ref{fig:m2Re500_spacetime}. Two regions where the instability grows at the source, around $x=20$ and $x=40$, can be observed as two vertical bars. This indicates that a global instability is present. The latter of those regions moreover seems to trigger strong convective instability bursts, \textit{i.e.} inclined lines which represent wavepackets traveling downstream through the domain with a front speed close to $1$. The nature of the growing global instability is revealed by looking at the streamwise velocity field at $t=3100$ (corresponding to the uppermost part of the space-time diagram) in figure \ref{fig:m2Re500_antisymmetric}, especially its antisymmetric component (bottom). This is a typical stationary Coanda-type global instability mode, which does not oscillate in time but simply deflects the jet from a symmetric position in the middle towards one of the walls. The symmetric jet has two recirculation bubbles placed symmetrically along each wall. As a result of the Coanda instability, one bubbles has grown in size and the one at the opposite wall has shrunk. The result is a new asymmetric steady state. However, the larger bubble may also trigger convective instability bursts; recirculation bubbles are known to exhibit strong convective instabilities~\cite{Marquet:2009p1247}. Similar bursts, "intermittency", developed around a Coanda-type asymmetric flow in a stenosis~\cite{johnstenosis}. 

By examining the green region in Fig.\ \ref{fig:Re_inner}, it appears that the Coanda instability occurs only for $m<0.9$, {\it i.e.} when the outer fluid is less viscous than the inner one. This is because the \textit{symmetric} base flows with lower outer fluid viscosity contain long regions of reverse flow. In~\cite{Xjunction}, a critical length of base flow recirculation zones ($L_r\approx 6$) was found at the onset of Coanda instability in a cross-junction, for several different parameters. For our jets at $Re=230$, the flow with uniform viscosity contains practically no reverse flow (it has minimum streamwise velocity $U=-10^{-4}$), and neither do the flow with higher viscosity outside. The length of the recirculation zones as a function of $m$ at $Re=230$ is shown in figure \ref{fig:Recircvsm}, top, blue line, which shows that the recirculation zones severely lengthen towards lower $m$ (up to $L_r=60$). The red line shows the critical recirculation length at the onset of the Coanda instability for these jets, which stays relatively constant between $L_r=6-10$ \footnote{The critical recirculation length varies more than in~\cite{Xjunction} probably because the instability in the present work is subcritical in some regions and supercritical in others}. The lengthening of the recirculation zone explains why the flow becomes more unstable at higher viscosity contrasts at the lower end (figure \ref{fig:Re_inner}). A visual demonstration of the separated flow is provided in figure \ref{fig:Recircvsm}, bottom, showing the streamwise velocity and its zero contour at $m=0.2$.

For $m<0.9$, the first bifurcation always happens through a stationary Coanda mode. For $0.7<m<0.9$, the Coanda instability is supercritical. For $m<0.7$, the Coanda instability is subcritical since a hysteresis is observed: at a fixed $Re$ and $m$, when starting from a symmetric base flow as an initial condition the final state is symmetric, but when starting from an asymmetric solution at higher Reynolds number as an initial condition, the final state is asymmetric. The boundary between blue and green regions denotes the subcritical instability boundary; on the left side (stable flow), the final flow state is symmetric irrespective of the initial condition. In~\cite{johnstenosis}, it was observed that a larger hysteresis region can be obtained when making analytic continuations with respect to other parameters than the Reynolds number. If this was done, the boundary may be pushed further to the left. However, in this study we focus on the nature of the instability at different viscosities, and therefore further parameter studies have been omitted. Results from all simulations are listed in the table in Appendix \ref{app:tab}, where the hysteresis ranges can be found.  

At Reynolds numbers above the first bifurcation, a possible mode competition between oscillatory and stationary global modes can be observed before arriving at the stationary asymmetric flow. This is indicated in figure \ref{fig:m0dot2Re500_withtime} at $m=0.2$, $Re=80$, where time increases from top to bottom. The instability starts in the form of high-frequency small-wavelength waves when the flow state is still nearly symmetric (top), but when the asymmetry develops (middle) these waves are slowly suppressed, until the flow arrives at a nearly steady state (bottom). This flow case belongs to the green region in figure \ref{fig:Re_inner}, where the final state is a steady and asymmetric jet. 

\begin{figure}
  \centering
  \includegraphics[width=\columnwidth]{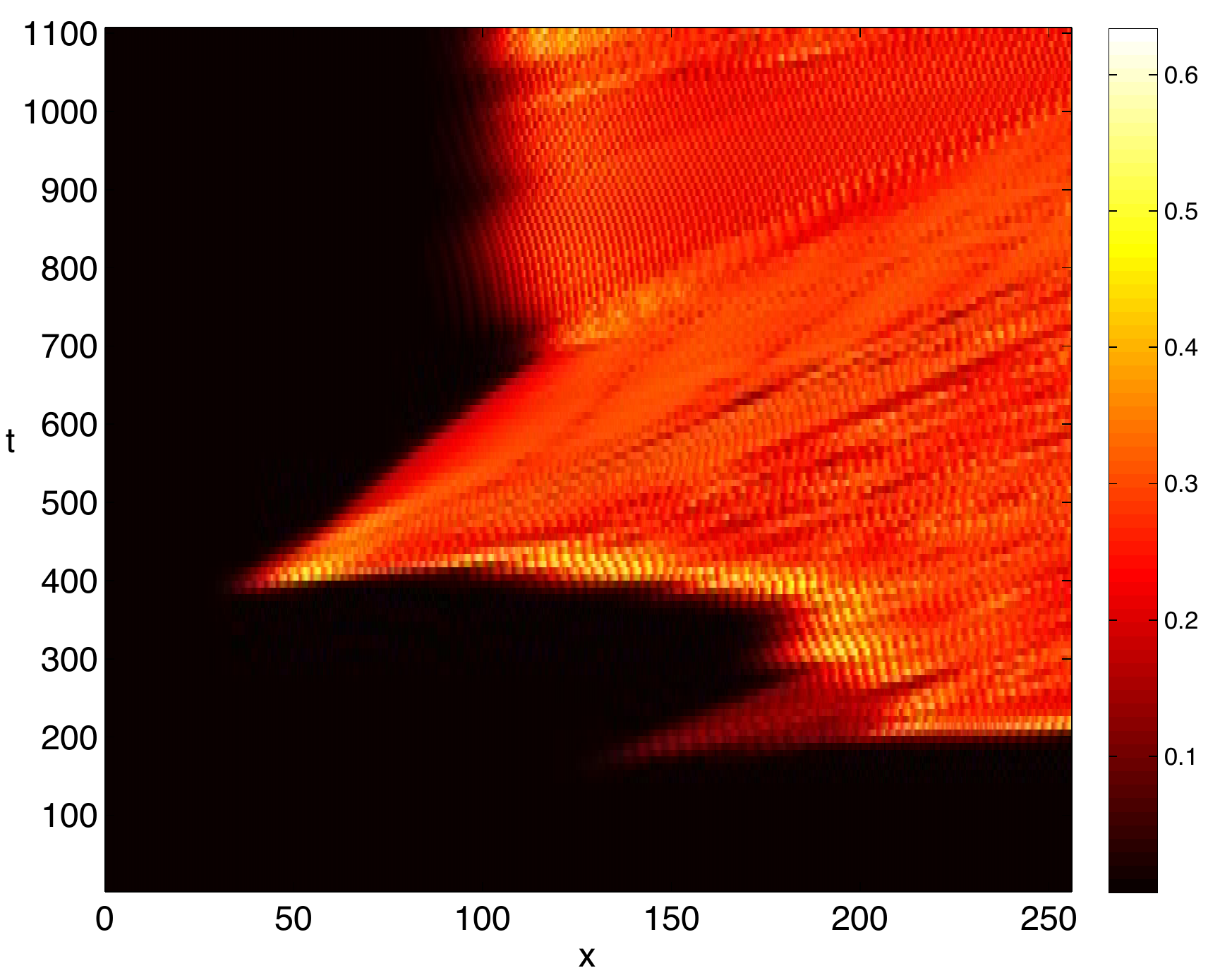}
  \caption{A space-time diagram of the kinetic energy of the antisymmetric perturbation along the lines $y=\pm1$, at $m=10$, $Re=250$. The downstream movement of the front in time shows that the instability occurs as convective bursts. This illustrates that the instability is convective (and not a global instability) for $m>1$.}
  \label{fig:m0dot1Re250_spacetime}
\end{figure}

\begin{figure}
  \centering
  \includegraphics[width=\columnwidth]{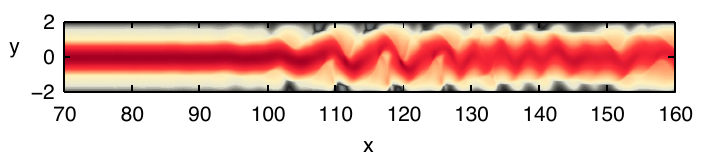}
  \includegraphics[width=\columnwidth]{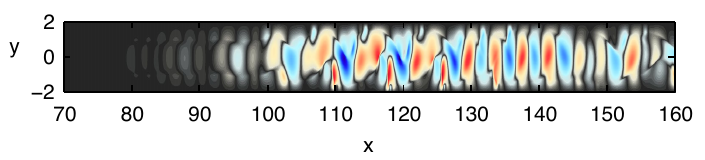}
  \caption{The saturated sinusoidal oscillation of the jet at $m=10$, $Re=250$: streamwise velocity (top), and vertical velocity (bottom). Note that the vertical axes in magnified by a factor $3$ compared to the horizontal axis.}
  \label{fig:m0dot1Re250_osc}
\end{figure}

\subsection{High viscosity of the outer fluid}
Now we move to the cases where the outer fluid is more viscous than the inner fluid. Also in this regime, a small viscosity contrast increases the critical Reynolds number (figure \ref{fig:Re_inner}) while for large viscosity contrast the critical Reynolds number decreases. The reason for destabilization at high viscosity outside is however very different from the destabilization at low viscosity outside (described in the previous subsection).

A space-time plot for $m=10$, $Re=250$, is shown in figure \ref{fig:m0dot1Re250_spacetime}. This shows oblique fronts of convective instability. The visible "front" finally settles at a location around $x\approx 100-120$, however the front is not stationary. The instability location is far downstream although this jet, being more viscous, reaches a fully developed profile very quickly. Both features indicate that the instability is convective. When looking at logarithmically spaced contours (not shown), the instability is seen to grow monotonously from upstream to downstream until the location where it reaches a visible amplitude and saturates. The shape of the final oscillation (at $t=1100$) is shown in figure \ref{fig:m0dot1Re250_osc}. This shows a sinusoidal oscillation of  large amplitude. A preliminary local stability analysis of these flows showed no absolute instabilities, but a very strong convective instability throughout the flow (with growth rates reaching O(1) upstream), and the sinusoidal mode had a higher growth rate than the varicose mode. This agrees with our hypothesis that the instability observed at $m>1$ is convective.It needs to be remembered that also strong enough convective instabilities in noise-amplifier flows may persist nonlinearly without triggering. However, the parameters at which they persist will strongly depend on the level of numerical/experimental noise (cmp. boundary layer instability and transition). 
\begin{figure}
  \centering
  \includegraphics[width=\columnwidth]{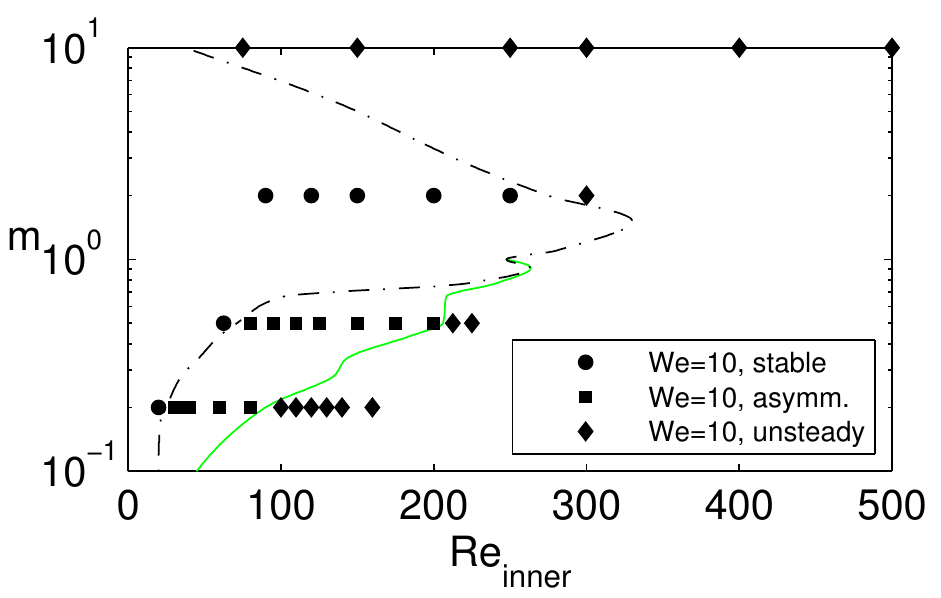}
  \includegraphics[width=\columnwidth]{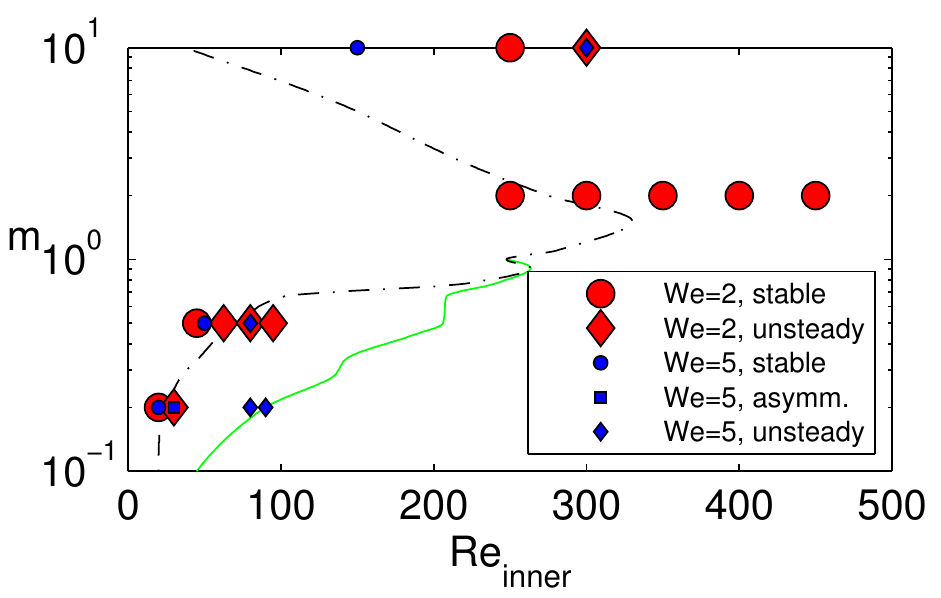}
   \includegraphics[width=\columnwidth]{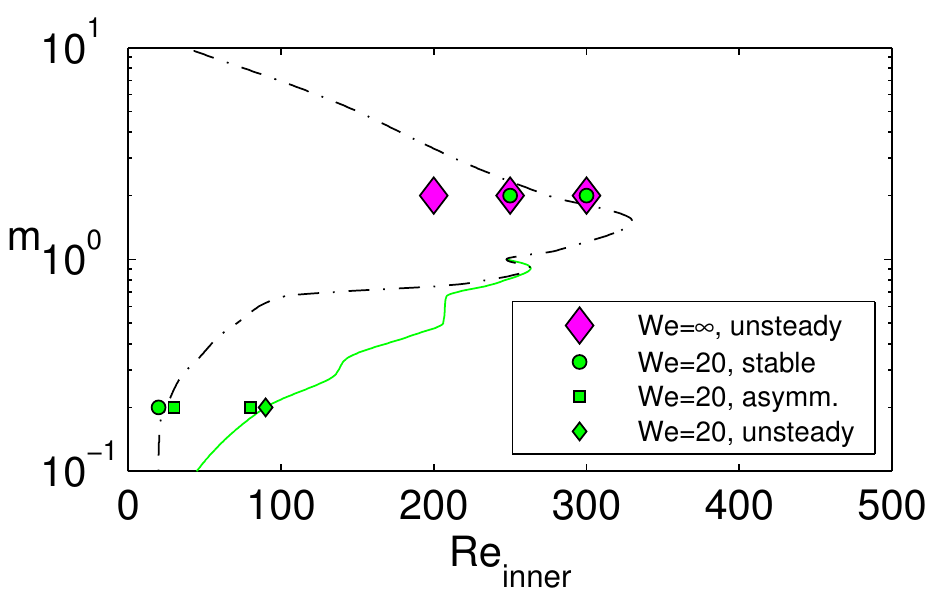}
  \caption{Results from selected simulations at varying $We$ (please see legend for exact values): $We=10$ (top, reference), lower $We$ (middle), and higher $We$ (bottom). The meaning of the markers: stable solution ($\circ$), steady asymmetric ($\square$), unsteady ($\Diamond$). Note that the unsteady solution is due to surface tension-induced mode at $We=2$, $m \leq 1$. 
  }
  \label{fig:Wevary}
\end{figure}
\begin{figure}
\centering
  \includegraphics[width=\columnwidth]{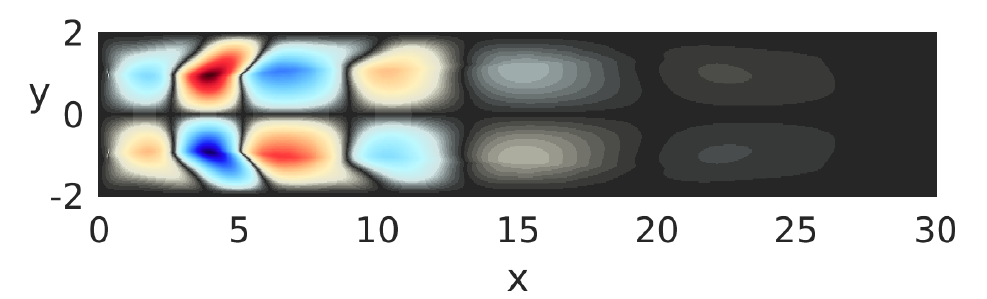}
   \includegraphics[width=\columnwidth]{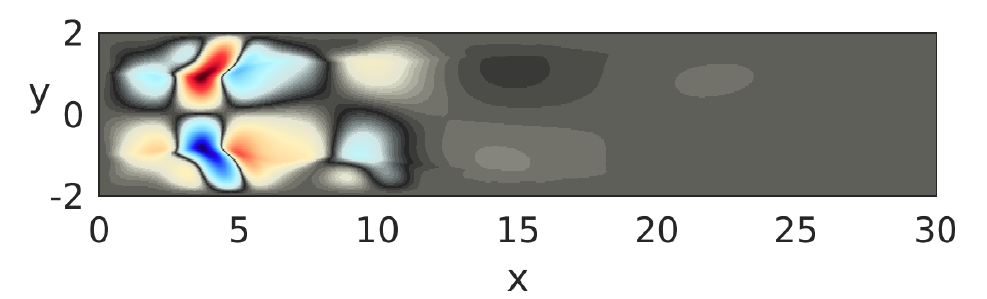}
   \includegraphics[width=\columnwidth]{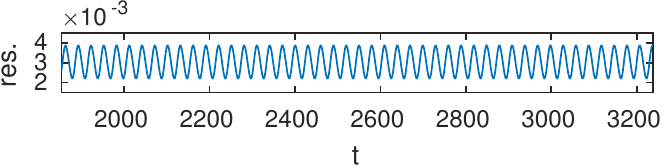}
  \caption{The first bifurcation by the surface tension-induced mode at $We=0.5$, low viscosity of the outer fluid: vertical velocity $m=0.5$ and $Re=62.5$ (top), vertical velocity $m=0.2$ and $Re=62.5$ (middle), saturated residual as a function of time for $m=0.5$ and $Re=80$ (bottom).
  }
  \label{fig:Wevary_surftensmode}
\end{figure}
\subsection{Influence of the Weber number and the shear ratio \label{sec:infl}}
All results so far were computed with a fixer Weber number ($We=10$). It is worth considering qualitatively how the instability mechanisms change when Weber number varies. To examine this, we have performed simulations at four selected viscosity ratios and four different Weber numbers. Figure \ref{fig:Wevary} (a) repeats the same results at $We=10$ for a visual reference. 
 
We have not observed completely new instability mechanisms when $We$ varies, but the neutral curves for each of the three modes described in the previous sections can move significantly. The most interesting effect is seen at low $We$ (high surface tension), at low viscosity of the outer fluid. Figure \ref{fig:Wevary} (b) shows the simulation results at low $We$. At $We=2$ (large markers, red online), the first bifurcation is directly time-dependent. This is seen from that the round markers (stable flow) are adjacent to diamonds (unsteady flow). Inspection of the mode shapes (Figure \ref{fig:Wevary_surftensmode} a--b) and the time signal (Figure \ref{fig:Wevary_surftensmode} c) reveals that the first bifurcation in this case is due to the surface tension-induced mode. The time signal is perfectly periodic, and there is no sign of asymmetry. Furthermore, the mode shape is very similar to the low Weber number symmetric instability modes for jets (and wakes) found in global instability analyses~\cite{VakarWe,StableWe}. The mode has a long wavelength and is localized close to the inlet.

At $We=2$, there is no indication of asymmetry (Coanda attachment), as there was at $We=10$ (Fig. \ref{fig:Wevary} a), where round markers were followed by squares. This means that the whole bifurcation sequence is altered at high surface tension, and suggests that there indeed is a mode competition when the viscosity of the outer fluid is low. At $We=10$, the growth of the Coanda attachment mode lead to an asymmetric flow, which suppressed the surface tension-induced mode. At $We=2$ on the other hand, the surface tension-induced mode is so strong that it prevents the Coanda mode from growing. The $m=0.5$, $Re=80$ case (Fig. \ref{fig:Wevary_surftensmode} a) was started from the asymmetric steady flow at $We=10$. Even in this case, the flow developed a surface tension-induced mode oscillating around a \textit{symmetric} mean.
At $We=5$ (slightly weaker surface tension), the first bifurcation is due to the surface tension-induced mode at $m=0.5$, but due to the Coanda attachment mode at $m=0.2$. At $We=20$ (much weaker surface tension), the first bifurcation is clearly due to the Coanda attachment mode. From this, we conlude that the instability region for the surface tension-induced mode grows constantly when $We$ decreases to low enough values, on the expense of the Coanda mode. This is logical, as Coanda attachment mode appears in single-phase flows, is related to the recirculation zones, and unrelated to surface tension. 

Let us now consider the flows with more viscous outer fluid. Figure \ref{fig:Wevary} (b) shows that at high values of surface tension ($We=2$, $We=5$), the flow is significantly stabilized at both $m=2$ and $m=10$. Correspondingly, figure \ref{fig:Wevary} (c) shows that without surface tension ($We=\infty$), the flow is significantly destabilized at $m=2$. Surface tension exerts the usual stabilizing influence on the oscillatory convective instability due to viscosity contrast. When the flow is initiated from an asymmetric steady solution, dissipative effects by surface tension are not important unless curvature is very large. Concluding, the surface tension-induced global instability is likely to stabilize for high and low Weber numbers~\cite{StableWe}. The other instability mechanisms remain creating two separate instability regions for high-viscosity jets and low-viscosity jets, respectively. The most unstable viscosity contrast will be at very low viscosity of the outer fluid due to Coanda attachment.
 
The influence of the shear ratio deserves a brief consideration. A detailed parameter study on the surface tension-induced global instability of uniform-viscosity jets was presented in~\cite{StableWe}. There it was shown that increasing shear ratio (a stronger co-flow at the inlet) was stabilizing for all Reynolds numbers studied ($Re \leq 500$). However, the largest shear ratio at which the flow was globally unstable increased with increasing Reynolds number. We conclude that increasing shear ratio should be always stabilizing for surface tension-induced mode. 
We also expect that the Coanda attachment mode is stabilized for high enough shear ratios, as a high co-flow is likely to eliminate the recirculation zones at the walls. The convective instability at high viscosity of the outer fluid is not inflectional, but appears even in channel and Couette flows as well~\cite{yih,hooperboyd1983,Boomkamp_IJMF_1996}. Hence, the instability at high viscosity of the outer fluid will not be qualitatively affected by changing shear ratio. In conclusion, increasing shear ratio would stabilize except at high viscosity of the outer fluid, and decreasing shear ratio would destabilize except at high viscosity of the outer fluid.
\begin{figure}
  \centering
  \includegraphics[width=\columnwidth]{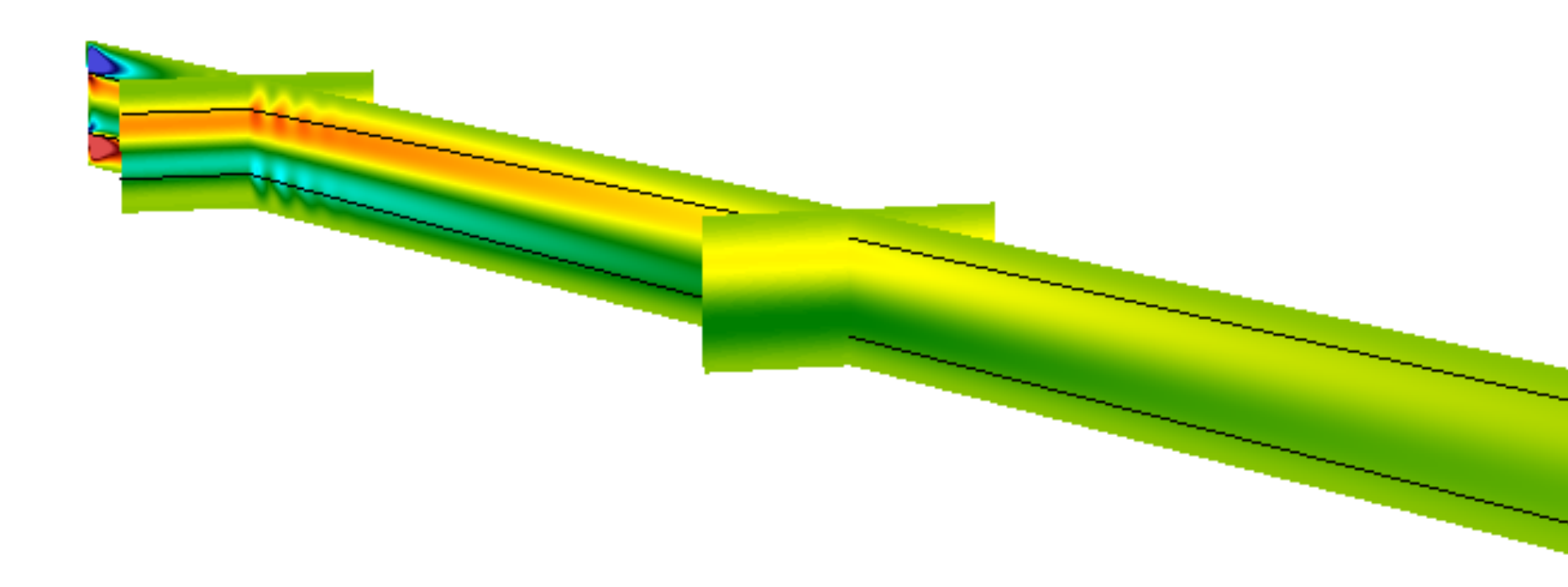}
  \caption{Instantaneous vertical velocity from a three-dimensional direct numerical simulation at $m=1$, $Re=250$, shown together with the interface position (black line) at selected cross-sections: $z=0$, $x=16$ and $x=64$. The spanwise slices ($x=16$, $x=64$) show that neither the velocity nor the interface position vary in the spanwise direction; the waves are fully two-dimensional.}
  \label{fig:3Dflowfield}
\end{figure}
\subsection{Absence of three-dimensionality}
We performed a selected number of three-dimensional direct numerical simulations, to investigate whether three-dimensional effects could influence the instability onset or development. These studies were done at $We=10$, for three different viscosity ratios: $m=0.2$, $m=1$ and $m=5$, and three different Reynolds numbers: $Re=125$, $Re=175$, $Re=250$.  The flow domain used was $[L_x, L_y, L_z] = [128, 4, 4]$ in the streamwise, vertical and spanwise directions (in the same order). The grid size was $N_x=6094$, $N_y=N_z=128$, resulting in 67 million grid points. The flow was run for several thousands of nondimensional time units in each case. 

The results showed without exceptions a two-dimensional flow field with a maximum spanwise velocity magnitude of $10^{-14}$. A representative flow field from a three-dimensional direct numerical simulation is shown in figure \ref{fig:3Dflowfield}. The spanwise cross-sections show that the flow field remains fully two-dimensional. These results imply we can exclude three-dimensional effects on the instability boundaries, and that secondary three-dimensional instabilities such as ligament formation are also unlikely in the investigated parameter regime. Previous studies have found ligament formation in viscosity-contrasted flows~\cite{ONaraigh_JFM_2014}, but typically at a higher Weber number ($We=10$) and a higher viscosity ratio ($m=30$). Here, the surface tension is probably too strong to allow ligament formation. For low viscosity of the outer fluid on the other hand, we might have expected a three-dimensional around the wall recirculation zones. However, it appears that the streamlines in our flows are not curved enough for centrifugal instabilties to form around the recirculation zones. 
\begin{figure}
  \centering
  \includegraphics[width=\columnwidth]{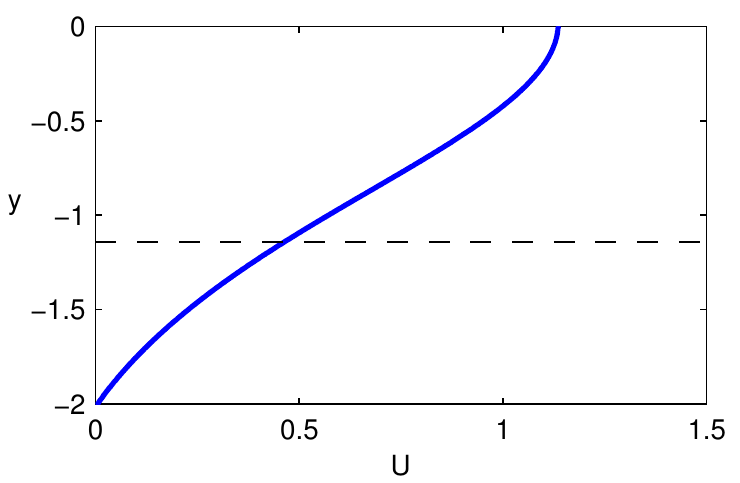}
  \includegraphics[width=\columnwidth]{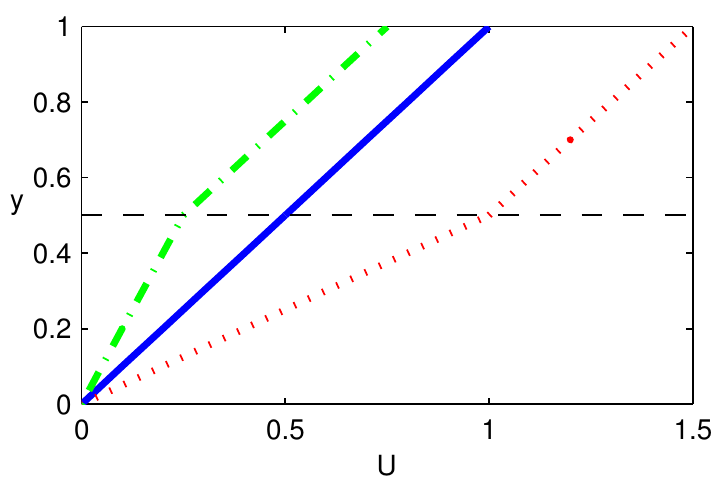}
  \caption{Top: Jet streamwise velocity profile from DNS at $m=1$, $Re=500$, $x=75$ (absolutely unstable due to surface tension), lower half. Bottom: Base flow profiles for the Couette model, at $m=1$ (solid, blue online), $m=0.5$ (dotted, red online) and $m=2$ (dash-dotted, green online). The problem is parameterized with respect to the shear of the upper layer.}
  \label{fig:couette_baseflowprofs}
\end{figure}
\section{Physical explanation for the appearance and disappearance of the surface tension-induced instability of jets \label{sec:local}}
It has been predicted~\citep{rees09} that surface tension may promote absolute instability in inviscid shear layers where Kelvin-Helmholtz instability is already present. The destabilization of wakes by surface tension has been previously explained by an inviscid mechanism using a broken-line shear layer profile~\cite{biancofiore_surftens_pof}. For wakes, the local and global instability is indeed located close to the inlet, where the velocity profile is strongly inflectional, and hence the destabilization of wakes could be explained by this model. However, jets have been observed to have convective instability due to surface tension for nearly-parabolic profiles~\cite{StableWe}. In the present work, also the absolute instability was seen to persist until $x>70$ (Fig.\ref{fig:m1Re500_omega0}), where the velocity gradient is nearly constant, as shown in figure \ref{fig:couette_baseflowprofs}. This suggests that the destabilization of jets could be due to another mechanism. The aim of the present section is to suggest a mechanism for the surface tension-induced instability of jets, and why it disappears when viscosity contrast is introduced. The aim is to give the simplest possible physical explanation for the neutral curve (fig. \ref{fig:Re_inner}). 

Absolute instability can be seen as the counterpart of global instability in parallel flows (unless the instability mechanism itself requires a non-parallel flow, such as recirculation). To become absolutely unstable, the flow also needs to be temporally unstable (or equivalently, convective instability precedes absolute instability). At uniform viscosity and in the absence of surface tension, exactly one branch of local temporally unstable modes (here called \textit{the jet mode}) exist for our jets. However, the jet mode never becomes absolutely unstable.  

In the presence of surface tension however, another branch of modes appears and becomes unstable, both convectively and absolutely. This second mode, \textit{the interfacial mode}~\cite{StableWe}, is a hidden mode in any flow with uniform density and viscosity, which appears as a neutral line in the spectrum only if interfacial perturbations are considered. Since the interface has no influence on the flow without surface tension, the neutral line thus corresponds to a pure convection of the interfacial perturbation by the local mean flow: $\partiel{h}{t}+U\partiel{h}{x}=0$. However, when even small surface tension or viscosity/density differences are introduced, the interface perturbation starts to interact with the flow, and loses its neutral stability. We should point out that this is the same mode which is responsible for the Yih instability~\cite{yih}. In that case, the interfacial mode described by Yih~\cite{yih} as "{\it a hidden neutral mode, ignored in conventional stability analyses}" locally destabilizes Couette flow for all Reynolds numbers with even a minor viscosity contrast.

For jets, it is the interfacial mode that is responsible for the global instability~\cite{StableWe}, and it was hypothesized in~\cite{StableWe} that the interaction of the two jet shear layers was not necessary but that the same instability could occur for a single shear layer. We now investigate this using the simplest possible model with the same main ingredients: shear, interfacial perturbation, viscosity gradient and confinement. The model chosen is a Couette flow with moving upper wall, and two fluid layers occupying half of the channel each. A second reason for chosing this model system was to confirm that no inflection points are needed for surface tension-induced instability: only shear and surface tension are needed for the flow to become locally and globally unstable, as was hypothesized in~\cite{VakarWe}. This model also covers the mechanisms for Yih instability and short-wave instability, and hence may shed light on what happens for viscosity-contrasted jets. 

The Reynolds and Weber numbers are based on the shear, and the channel height. The upper layer has more momentum and hence plays a role similar to the jet inner flow. The nondimensionalisation is thus based on the parameters of the upper layer: $Re=(dU/dy)_{up}*H^2$, $Re=(dU/dy)_{up}*H^2$.  Note moreover that, for $m=1$, the velocity gradient for both layers is the same. Representative base flow profiles are shown in figure \ref{fig:couette_baseflowprofs}. The Ansatz for the velocity, pressure and interfacial perturbation is of the form:
\begin{equation}{\textbf{U}_{tot}=\textbf{U}_b (y)+\hat{\textbf{u}}(x,y) \exp\left\{\textrm{i}\alpha x+\sigma t\right\},}\end{equation} 
\begin{equation}{P_{tot}=p(x,y) \exp\left\{\textrm{i}\alpha x+\sigma t\right\},}\end{equation} 
\begin{equation}{h_{tot}=0.5+\hat h \exp\left\{\textrm{i}\alpha x+\sigma t\right\},}\end{equation} 
where $\textbf{U}$ is the base flow velocity field, $\hat u(x,y)$ is the spatial shape of the velocity eigenmode, $\sigma=\sigma_r+\textrm{i}\sigma_i$ is the temporal eigenvalue ($\sigma_r>0$ unstable), and $\alpha$ is the streamwise wavenumber. For these studies, we solve the same equations as in~\cite{StableWe} using the FLUIDSPACK code, with the disturbance $x$-derivatives replaced by $\textrm{i}\alpha$ and base flow $x$-derivatives set to zero.

\begin{figure}
  \centering  
  \includegraphics[width=\columnwidth]{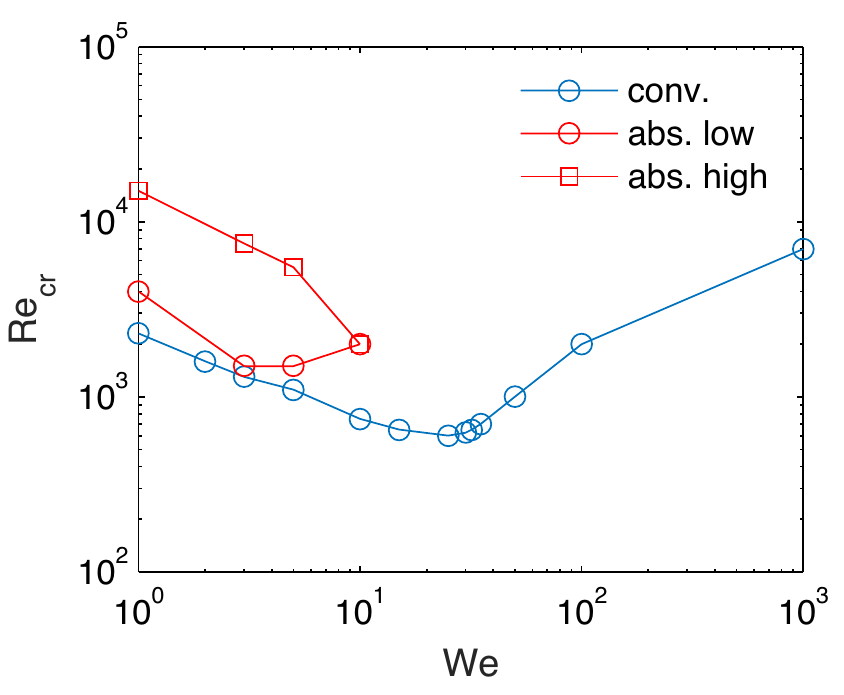}
  \caption{Critical Reynolds number for the onset of instability (convective and absolute) with surface tension, Couette flow model. The upper limit for absolute instability is also depicted (convective instability exists for all finite $Re>Re_{cr}$).}
  \label{fig:Couette_Recr_m1}
\end{figure}

\begin{figure}
  \centering  
  \includegraphics[width=\columnwidth]{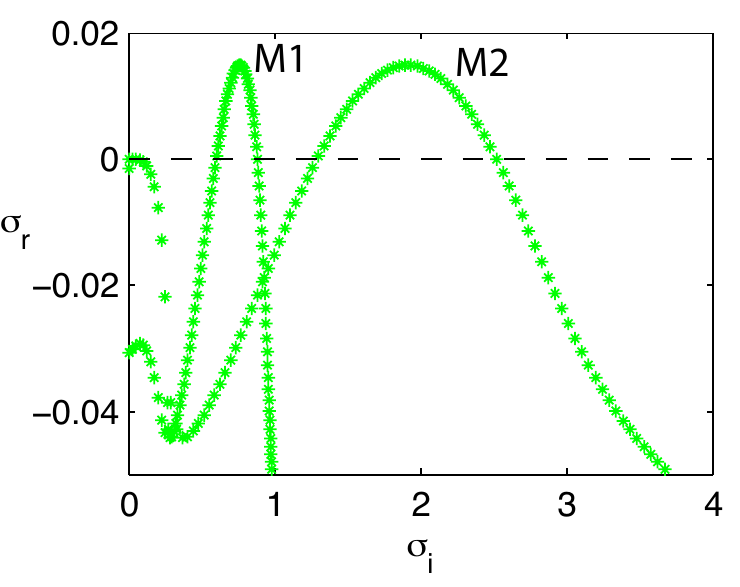}
  \caption{Eigenvalue spectrum (all wavenumbers) for the two-fluid Couette flow model, at $m=1$, $We=25$ and $Re=1300$. Two unstable modes are seen.}
  \label{fig:twomodes_couettem1}
\end{figure}

\begin{figure}
  \centering  
  \includegraphics[width=\columnwidth]{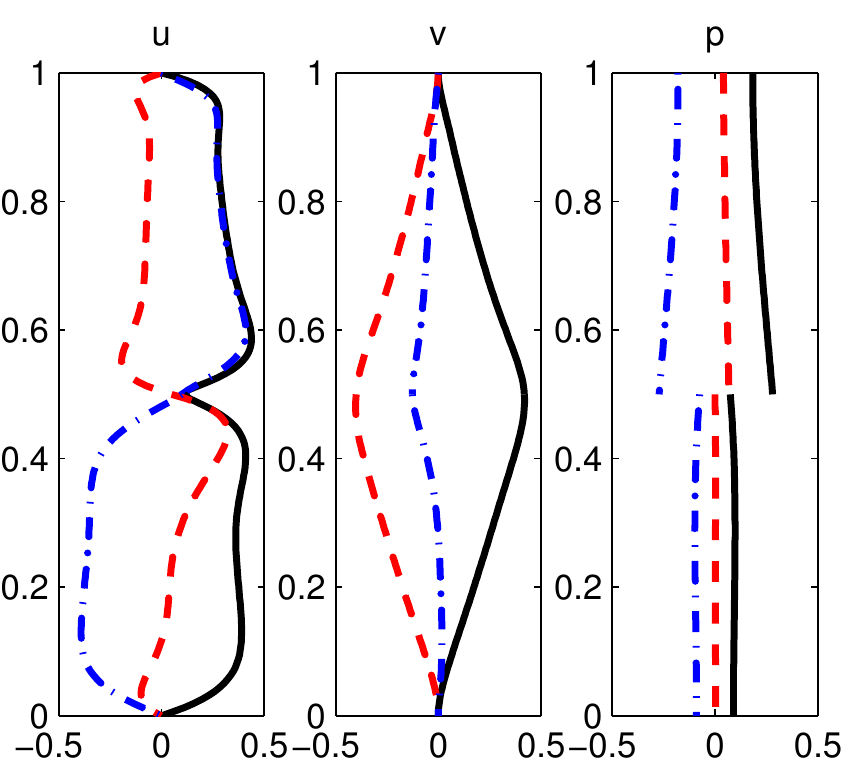}
  \caption{Eigenfunction for the slower unstable (interfacial) mode.}
  \label{fig:eigfun1_reimag}
\end{figure}

\begin{figure}
  \centering  
  \includegraphics[width=\columnwidth]{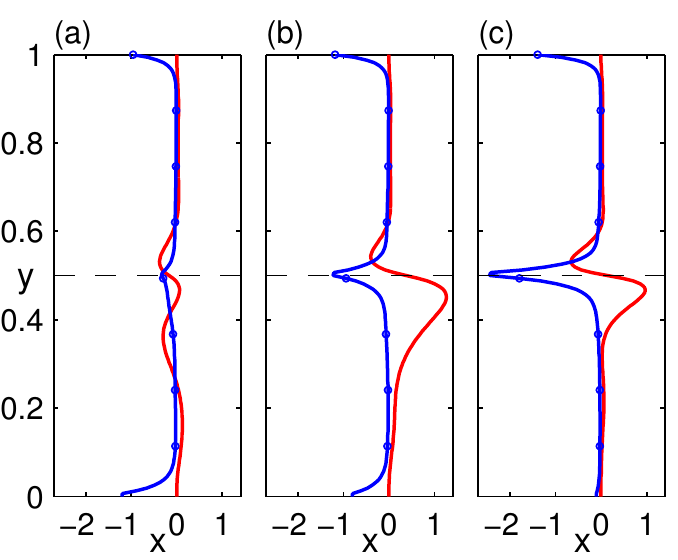}
  \caption{Vertical distribution of the production (solid, red online) and dissipation (dash-dot, blue online) of kinetic energy for the eigenmode at $m=1$, $Re=1300$ and $\alpha_r=2.7$: (a) $We=100$ ($\sigma_r=-0.05$), (b) $We=25$ ($\sigma_r=0.02$), and (c) $We=10$ ($\sigma_r=-0.02$). The surface position is shown by a dashed horizontal line (black online).}
  \label{fig:proddiss}
\end{figure}

\begin{figure}
  \centering  
  \includegraphics[width=\columnwidth]{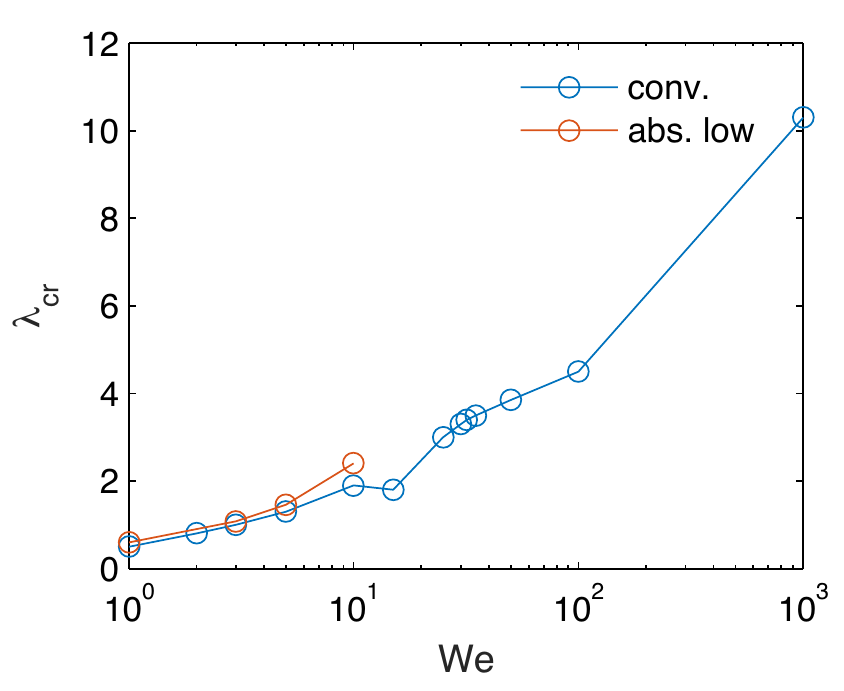}
  \caption{Wavenumber at the onset of instability (convective and absolute) with surface tension, Couette flow model. }
  \label{fig:Couette_lambdacr}
\end{figure}

\subsection{Destabilization by surface tension}

For our jets, surface tension destabilizes a hidden interfacial mode both locally and globally. The Couette flow model without Kelvin-Helmholtz instability reproduces the latter behavior. The critical shear-based Reynolds numbers for the onset of convective and absolute instability are shown in figure \ref{fig:Couette_Recr_m1} for varying Weber numbers. 

Let us first analyse the convective instability. The local temporal (\textit{i.e.} convective) instability spectrum for an unstable case ($We=25$, $Re=1300$, $m=1$) is shown in figure \ref{fig:twomodes_couettem1} for varying $\alpha_r$. Two unstable modes appear, with exactly the same growth rate for the same $\alpha_r$, but with different phase speeds. The eigenmode shape of the slower mode (M1) at the wavelength corresponding to instability maximum ($\alpha_r=2.7$) is shown in figure \ref{fig:eigfun1_reimag}. The kinetic energy of the M1 mode is symmetrically distributed above and below the interface (at $y=0.5$). The faster mode (M2) has a similar shape but its amplitude maximum is located in the faster fluid, and due to this it seems to have a higher group velocity and never becomes absolutely unstable. Here, we focus on the slower mode (M1), because it also has the lowest group velocity, and is the one which becomes absolutely unstable.

Mode M1 is present for all Reynolds numbers above a critical threshold. We have analysed values up to $Re=500 \ 000$. However, the growth rate decays towards zero for high Reynolds numbers. Consequently, surface tension destabilizes the interfacial mode by a viscous mechanism. We can analyse this further by separating the components which cause growth of the perturbation kinetic energy. The normalized kinetic energy of an eigenmode $E_{kin}^{-1}(dE_{kin}/dt)$ grows or decays at the same rate as the mode. Hence the components of this expression can be used to analyse how different mechanisms contribute to the eigenvalue growth. The perturbation kinetic energy equation (derived in Appendix \ref{app:ener}) becomes:
\begin{eqnarray}
\frac{d E_{kin}}{dt}&=& \nonumber \\
\int^{(1),(2)} \partiel{||\textbf{u}||^2}{t}&=& \nonumber \\
\int^{(1),(2)} -\left(u^*v+v^*u\right)U'&&\nonumber \\
+\int^{(1)} -\frac{2}{Re} \left(||\alpha \textbf{u}||^2 +||D\textbf{u}||^2\right)&&\nonumber \\
+\int^{(2)} -\frac{2m}{Re} \left(||\alpha \textbf{u}||^2 +||D\textbf{u}||^2\right)&&\nonumber \\
+ W_{We} + W_{m} &&
\end{eqnarray}
where the work performed by the surface tension is:
\begin{equation}{W_{We}=\int_B -\frac{\alpha^2}{We}\partiel{||h||^2}{t}}\end{equation}
and the work performed by viscosity contrast at the interface is:
\begin{equation}{W_{m}=\int_B \frac{(m-1)}{m Re}\left[2\alpha^2 U ||h||^2+Du^{(1)}h^*+Du^{(1)*} h \right]}\end{equation}
The sign of each integral component tells us whether it is stabilizing or destabilizing, and their relative magnitudes can be compared as well. The first volumetric integral is the well-known kinetic energy production by base flow shear, and the second and third integral are the viscous dissipation in each domain. The surface term $W_{We}$ is the energy dissipation due to surface tension. This term is negative whenever the eigenmode growth is positive; it is known that surface tension in itself cannot destabilize parallel flows. The surface term $W_m$ is the energy production due to a viscosity contrast, which is zero for $m=1$. This means that the surface tension-induced (temporal) instability must be due to the volumetric terms: the energy production and dissipation inside the domain. It is known that inviscid flows cannot have kinetic energy production (from the base flow shear) without inflection point, and hence it is only logical that the surface tension-induced instability is viscous.

The magnitude of the three terms can be compared to each other for M1. For the unstable cases, production and dissipation are both larger than the term due to surface tension. For instance at $We=25$ (the most unstable Weber number in Fig.\ \ref{fig:Couette_Recr_m1}) at $Re=1300$ and $\alpha=2.7$ (the most unstable wavenumber for this case), the magnitudes are $0.18$ for production, $-0.13$ for viscous dissipation, and $-0.026$ for dissipation by surface tension. The vertical distribution of production and dissipation is shown in figure \ref{fig:proddiss}, for $We=100$ (a), $We=25$ (b), and $We=10$ (c). This shows that at low surface tension (a), there is not much production neither dissipation. When surface tension increases to $We=25$ (b), there is efficient energy production in the slower fluid, and some dissipation at the surface. When the surface tension increases further (c), the production decreases while the dissipation near the surface increases. In conclusion, the right amount of surface tension destabilizes the M1 mode by regulating the delicate balance between production and dissipation, and the dissipation by surface tension itself is negligible in comparison to this effect. 

Based on these figures, the instability mechanism might be hypothesized as follows: A wave-like perturbation of the interface induces a wave-like perturbation of pressure, in order to satisfy the stress balance in the presence of surface tension. The streamwise pressure gradients in turn induce a streamwise velocity perturbation, and vertical velocity perturbation by mass conservation. This is the flow which pulls the interface back to its flat position. Thinking about a stationary situation, a perturbation which pulls the interface back must be symmetric with respect to the interface for the vertical, and antisymmetric for the streamwise velocity. Figure \ref{fig:eigfun1_reimag} shows that M1 has this symmetry property, and Fig.\ref{fig:Wmode_m1_growing} shows clearly that the jet global mode has this symmetry as well. This tendency to pull the interface back results in a situation where $u$ and $v$ have opposite signs in the slower fluid, which makes it possible to extract energy from the mean shear in the slower fluid. The presence of viscosity makes it possible to extract energy through the term $-uvU'$ even without inflection points; however, near the surface the mode experiences high level of viscous dissipation at the surface because of the steep gradient of the mode du/dy. Surface tension dictates the vertical scale of M1 --- a high surface tension focuses the mode close to the surface resulting in high viscous dissipation near the surface (Fig. \ref{fig:proddiss} c), while a low surface tension results in a weaker mode with neither production nor dissipation (Fig. \ref{fig:proddiss} c). Intermediate surface tension (Fig. \ref{fig:proddiss} b) allows the mode to penetrate deep enough in the lower fluid to have efficient production while keeping the surface gradients moderate.

Finally, surface tension introduces damping and hence slows down the oscillation frequency. As waves of high wavenumbers have higher curvature, they seem to be slowed down by surface tension more than low wavenumbers. Near capillary wavelength, the change in phase speed $c=\omega/k$ is particularly rapid, and this may result in unstable waves with zero group speed. In inviscid Kelvin-Helmholtz shear layer~\cite{rees09}, surface tension induced absolute instability. Here, we show that in the Couette model system without inflection points, surface tension induces an absolute instability of M1 (the red lines in figure \ref{fig:Couette_Recr_m1}). This happens only when surface tension is strong enough, at low Weber numbers ($We<15$). The critical (lowest) Reynolds number at which the absolute instability appears is a function of the Weber number. However, there is also a highest Reynolds number above which the absolute instability disappears. 

Extrapolating this information from M1 to the observed jet mode with the same structure, we hypothesize that the global surface-tension-driven instability is viscous and disappears when the Reynolds number is increased. Finally, the wavenumber at the onset of convective and absolute instability is depicted in figure \ref{fig:Couette_lambdacr}. The convective and absolute instabilities of the interfacial mode appear at similar wavenumbers, and the wavenumber $\alpha_{cr}$ at the instability onset increases with Weber number.

\begin{figure}
  \centering  
  \includegraphics[width=\columnwidth]{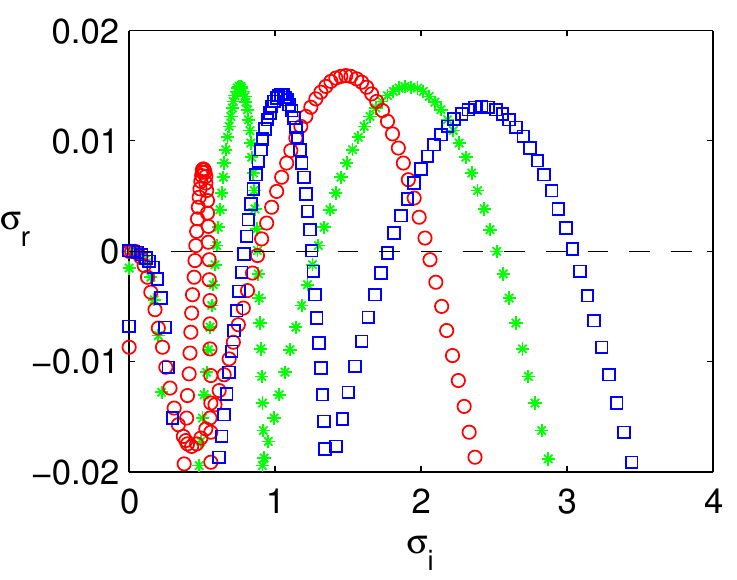}
  \caption{Eigenvalue spectrum (all wavenumbers) for the two-fluid Couette flow model, at different viscosity ratios, $We=25$ and $Re=1300$: stars (green online) $m=1$ (uniform viscosity), circles (red online) $m=1.3$ (high viscosity in the slower fluid), squares (blue online) $m=0.8$ (low viscosity in the slower fluid).}
  \label{fig:twomodes_couetteviscvar}
\end{figure}

\begin{figure}
  \centering  
  \includegraphics[width=\columnwidth]{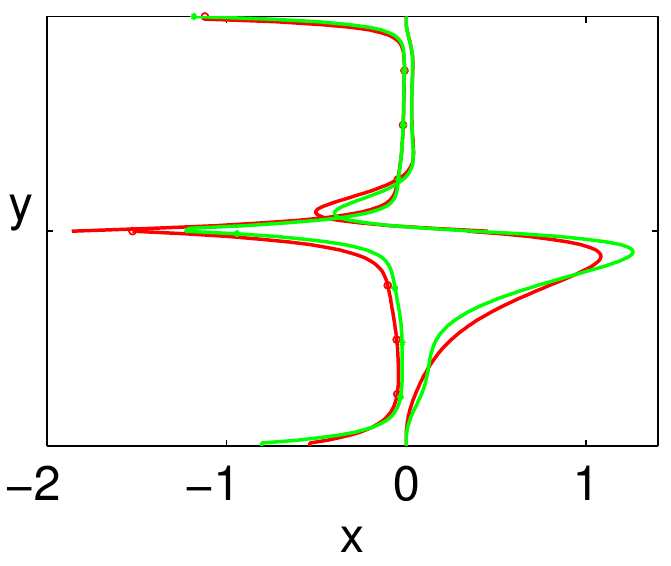}
  \caption{Production (plane solid line) and dissipation (line with markers) distributions, as in figure \ref{fig:proddiss}, for the interfacial mode at $m=1$ (green online) and for $m=1.3$ (red online).}
  \label{fig:proddiss_couetteviscvar}
\end{figure}

\begin{figure}
  \centering  
  \includegraphics[width=\columnwidth]{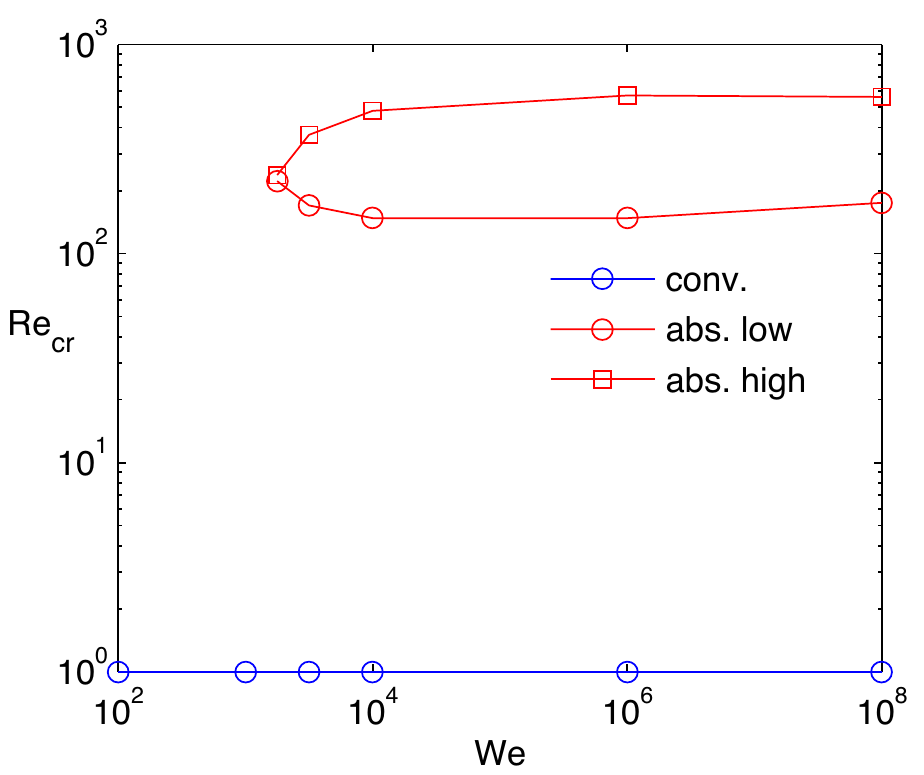}
  \caption{The same as figure \ref{fig:Couette_Recr_m1} but for viscous outer fluid: $m=10$. The flow is convectively unstable for all $Re$ through the Yih mechanism. ($Re=1$ was the lowest Reynolds number tested.)}
  \label{fig:m10couette}
\end{figure}

\subsection{Effect of a viscosity ratio}

Now, a viscosity ratio is introduced while keeping the shear-based Reynolds number of the upper layer constant. 
Figure \ref{fig:twomodes_couetteviscvar} shows how the frequency and growth of the two interfacial modes (M1 and M2) changes at $We=25$, $Re=1300$, when a viscosity ratio is introduced. The green line shows the $m=1$ case. 

The blue markers in Fig. \ref{fig:twomodes_couetteviscvar} show M1 and M2 with a less viscous outer fluid ($m=0.8$). The growth of M1 is slightly smaller than for $m=1$. The main change, however, is that the phase velocity of M1 (and M2) increases considerably. This is because M1 travels at the surface velocity, and the surface velocity increases to satisfy the base flow stress balance(Fig. \ref{fig:couette_baseflowprofs}). This increase of the convection velocity happens also for the jet. It turns out that this increases the group velocity of the mode. Absolute instability is weakened, and seems to disappear for viscosity ratios $m<0.7$. In the DNS, we observed an unstable stationary Coanda global mode. However, the Coanda mode is not due to local absolute instability, but requires an essentially non-parallel flow (it strongly depends on the size of a closed recirculation zone). Hence, we cannot observe Coanda instability in the Couette model. However, the Couette model predicts that the surface tension-induced instability should stabilize at low outer fluid viscosity, which agrees with our observations from the DNS.

The red markers in Fig. \ref{fig:twomodes_couetteviscvar} show M1 and M2 with a more viscous outer fluid ($m=1.3$). Now, both modes have slowed down because the interfacial velocity has decreased. However, the growth rate of M1 has halved. We have done an energy analysis at the point of maximum growth of M1, which is still at $\alpha=2.7$. This reveals that the energy production by base flow shear ($0.17$) is slightly smaller than in the uniform viscosity case, and there is more viscous dissipation: $-0.16$. The energy dissipation due to surface tension is $-0.014$, and the energy production by viscosity contrast $0.014$. Hence, the surface terms take out each other, and their contributions are an order of magnitude smaller compared to the volumetric terms; the stability is regulated by the balance between production by the base flow shear and viscous dissipation. The production and dissipation distributions are shown in Fig.\ \ref{fig:proddiss_couetteviscvar}, illustrating that viscous dissipation increases for $m>1$ both near the surface and in the more viscous fluid. 
The decrease of the temporal growth of M1 also pushes the zero group velocity point to the stable half plane; for $m>1$, the absolute instability of M1 soon disappears. It is interesting to note that the growth rate of M2 actually increases. However, M2 does not become absolutely unstable because of its high convection speed. This could explain the strong convective instability of the jets at $m>1$.

Summarizing, the viscosity ratio differs from unity, to either direction, the surface tension-driven absolute instability soon disappears. Most cases with lower viscosity of the outer fluid do not have any absolute instability. The cases with high viscosity outside shows the neutral curve display a very strong convective instability. The convective and absolute instability for $m=10$ is shown in figure \ref{fig:m10couette}. The flow is convectively unstable for all Reynolds numbers as expected by the Yih and shortwave instability mechanisms (the lowest Reynolds number tried here was $Re=1$). The case $m=10$ depicts some absolute instability, but with the opposite trend compared to surface tension-driven instability --- the absolute instability appears only at very high Weber numbers. This shows that the absolute instability mechanism is viscosity-driven, stabilized by surface tension. Extrapolating this information to our jet, the shear-based Weber numbers are so high that our jet at $m=10$ is not likely to have any absolute instability. However, at even higher viscosity ratios, the absolute instability is moved to lower Weber numbers in this Couette model. At $m=30$, the Couette flow is absolutely unstable at $Re=100$, similarly to the viscous liquid layer by~\cite{ONaraigh_JFM_2014}. Hence, while the high-viscosity cases presented in this paper are convective, if the viscosity ratio of the jet is increased even more, we are likely to observe another mode of global instability.

\section{Conclusions}
In this study, we have investigated how the viscosity ratio influences the global instability of an immiscible confined co-flow jet, by direct numerical simulations of the two-fluid system using a level-set method. The main focus has been to quantify how the surface tension-induced instability of jets is influenced by a viscosity contrast. We find that a small viscosity contrast in both directions is stabilizing, while at larger viscosity contrasts the critical Reynolds number is decreased due to other instability mechanisms which take over at high viscosity contrast. The instability mechanisms are further analysed using time-space diagrams.

For a co-flow of less viscous fluid outside the jet, when the Reynolds number is increased from zero, the first bifurcation is through a Coanda attachment instability, which makes the jet steady but asymmetric. This is because recirculation zones develop in the less viscous outer phase, and the length of these zones increases as a function of viscosity contrast, further decreasing the critical Reynolds number. When Reynolds number is increased, time-dependent convective bursts develop particularly around the larger recirculation zone. 
For a co-flow of more viscous fluid outside the jet, the first bifurcation is through a very strong convective instability which appears in the unforced DNS.

Finally, we analyse the origin of the local and global destabilization of a "hidden" interfacial mode in the jet flow by means of local stability analysis of a two-layer Couette flow model system. We show that a qualitatively similar local and global destabilization by surface tension is obtained for Couette flow, and that the mode is viscous (vanishes in inviscid flow). We show that the absolute instability occurs at uniform viscosity but vanishes with lower viscosity of the outer flow (due to increase of interfacial convection speed) and with higher viscosity of the outer flow (the mode extracts its energy in the slower fluid).
Finally, the viscous instability mechanism found here occurs without inflection points, and is different from the inviscid mechanism found by~ \cite{biancofiore_surftens_pof} in immiscible wake flows (by interaction between a Rossby wave at a vorticity discontinuity and a capillary wave). The viscous mechanism could explain the destabilization of these jets, particularly as nearly parabolic~\cite{StableWe} and constant-gradient jet profiles were destabilized in this work. However, both mechanisms can exist or co-exist for jets and wakes at different parameter regimes. 
\section*{Acknowledgements}
The work of O. Tammisola has received the financial support of Swedish Research Council (VR) through the Grant Nr. 2013-5789. J.-C. Loiseau and L. Brandt acknowledge support by the European Research Council Grant No. ERC-2013-CoG-616186, TRITOS and by the Swedish Research Council (VR). The authors acknowledge computer time provided by the Swedish National Infrastructure for Computing (SNIC) at PDC Centre for High Performance Computing (PDC-HPC). The authors also wish to thank Lennon O'Naraigh for introducing them to the TPLS solver. 
\bibliography{../../../../Rapporter/postdocref}
\appendix

\section{Derivation of the perturbation kinetic energy growth \label{app:ener}}
For an unstable local linear eigenmode, the kinetic energy and the interfacial energy both grow at the same (exponential) rate. Hence, to find mechanisms of eigenvalue growth we may choose to focus on the kinetic energy, because $\sigma_r=(1/E_{kin})\frac{d}{dt}(E_{kin})$.
\begin{equation}{
\begin{array}{rcl}
\int u_i\partiel{u_i}{t}+\int u_i U_j \partiel{u_i}{x_j}+\int u_i u_j \partiel{U_i}{x_j}+\int u_i u_j \partiel{u_i}{x_j}&=& \\
\int u_i\partiel{\tau_{ij}}{x_j},
\end{array}
}\end{equation}
where $\tau_{ij}=-p\delta_{ij}+(1/\tilde Re)\left(\partiel{u_i}{x_j}+\partiel{u_j}{x_i}\right)$ and $\tilde Re=Re$ in region 1 and $\tilde Re=Re/m$ in region 2.
This can be partially integrated \textit{at each of the two regions separated by the steady positions of the interface} to yield:

\begin{eqnarray}
\int \frac{1}{2}\partiel{(u_i u_i)}{t}+\int \frac{1}{2} \partiel{(u_i u_i U_j+u_i u_i u_j)}{x_j}+\int u_i u_j \partiel{U_i}{x_j}&=& \nonumber \\
\int \partiel{}{x_j}\left[u_i\left(-p \delta_{ij}+\tilde Re^{-1}\left(\partiel{u_i}{x_j}+\partiel{u_j}{x_i}\right)\right)\right] && \nonumber \\ 
+\int p \partiel{u_i}{x_i} -\int \partiel{u_i}{x_j}\tilde Re^{-1} \left(\partiel{u_i}{x_j}+\partiel{u_j}{x_i}\right) &=&\nonumber \\
\int_B \left[u_i \tau_{ij} N_j\right] -\int \tilde Re^{-1} \partiel{u_i}{x_j} \partiel{u_i}{x_j} && \nonumber \\
-\int \partiel{}{x_j} \left[\tilde Re^{-1}u_i \partiel{u_j}{x_i}\right]+\int \tilde Re^{-1} u_i \frac{\partial^2{u_j}}{\partial x_i \partial x_j} &=&\nonumber \\
-\int \tilde Re^{-1} \partiel{u_i}{x_j} \partiel{u_i}{x_j}+ \int_B \left[u_i \tau_{ij} N_j\right] -\int_B \left[\tilde Re^{-1} u_i \partiel{u_j}{x_i} N_j\right]  &&\nonumber \\
\end{eqnarray}
where $N_j$ is the outward pointing normal of the \textit{steady interface}, and several terms were eliminated by applying the continuity equation for the base flow and the perturbation.
Hence, this gives the following equation for the evolution of perturbation kinetic energy in \textit{each of the two regions}:
\begin{eqnarray}
\int \frac{1}{2}\partiel{(u_i u_i)}{t}&=&\int -u_i u_j \partiel{U_i}{x_j}+\int -\tilde Re^{-1} \partiel{u_i}{x_j} \partiel{u_i}{x_j}\nonumber \\
&&+\int_B \left[ -\frac{1}{2} (u_i u_i U_j+u_i u_i u_j)N_j\right]  \nonumber \\
&&+\int_B \left[u_i \tau_{ij} N_j\right] +\int_B \left[-\tilde Re^{-1} u_i \partiel{u_j}{x_i} N_j\right] \nonumber \\
\end{eqnarray}
where $N_i$ is the outward pointing normal from each domain. The first term is the well-known energy production in shear flows, the second dissipation due to viscosity (which is always negative). The rest are transport terms, which vanish at outer boundaries, but not necessarily at the interface between the two fluids.

In the Couette flow problem, we have: $\textbf{U}^{(1)}=(y+(m-1)/(2m),0)$ in the upper domain, and $\textbf{U}^{(2)}=(y/m,0)$ in the lower domain, which gives $U^{'(1)}=1$ and $U^{'(2)}=1/m$, $N^{(2)}=(0,-1)$, and $N^{(1)}=(0,1)$. 
Summing these up, the total rate of change of kinetic energy in the (Couette flow) system becomes:
\begin{eqnarray}
\int^{(1),(2)} \frac{1}{2}\partiel{(u u+v v)}{t}&=& \nonumber \\
\int^{(1),(2)} -uv U'&&\nonumber \\
+\int^{(1)} -\frac{1}{Re} \left(\partiel{u}{x}\partiel{u}{x}+\partiel{u}{y}\partiel{u}{y}+\partiel{v}{x}\partiel{v}{x}+\partiel{v}{y}\partiel{v}{y}\right)&&\nonumber \\
+\int^{(2)} -\frac{m}{Re} \left(\partiel{u}{x} \partiel{u}{x}+\partiel{u}{y}\partiel{u}{y}+\partiel{v}{x}\partiel{v}{x}+\partiel{v}{y}\partiel{v}{y}\right)&&\nonumber \\
-\int_B \Delta\left[\frac{1}{2} (u u+v v)v\right] &&\nonumber \\
+\int_B \Delta \left[u \tau_{12}+v\tau_{22}\right] &&\nonumber \\
+\int_B \left[-\frac{m}{Re}\left(u^{(2)} \partiel{v^{(2)}}{x}+v^{(2)} \partiel{v^{(2)}}{y}\right) \right. &&\nonumber \\
+\left.\frac{1}{Re}\left(u^{(1)} \partiel{v^{(1)}}{x}+v^{(1)} \partiel{v^{(1)}}{y}\right) \right] &=&\nonumber \\
\end{eqnarray}
We are interested in the time-averaged kinetic energy growth of a local stability eigenmode over a period $T=(2\pi/\omega)$ and a wavelength $\lambda=(2\pi/\alpha)$ .
From now on, all integrals should be interpreted as such averages. The complex temporal modal ansatz for the local analysis is:
\begin{eqnarray}
u(x,y,t)&=&\Re \left\{\hat u(y) \exp\left(\mathbf{i}\alpha x+\sigma t\right)\right\} \nonumber \\
&=&(1/2)\left(\hat u(y) \exp\left(\mathbf{i}\alpha x+\sigma t\right)+ \right. \nonumber \\ 
&& \left. \hat u(y)^*\exp\left(-\mathbf{i}\alpha x+\sigma^* t\right)\right), \nonumber \\
\end{eqnarray}
and similarly for other variables. The real quantities can be obtained by the transformation: $\Re \{u\}=(1/2)(u+u^*)$, where $^*$ denotes complex conjugate. 
The average over $T$ and $\lambda$ means that only the products of one non-conjugated and one conjugate variable survive the integration, whereas products of odd number of perturbation variables do not. This means that the first boundary term will cancel. Furthermore, the boundary terms containing $Re$ will cancel due to continuity. Finally, we obtain:    
\begin{eqnarray}
\int^{(1),(2)} \partiel{(u^*u+v^*v)}{t}&=& \nonumber \\
\int^{(1),(2)} -\left(u^*v+v^*u\right)U'&&\nonumber \\
+\int^{(1)} -\frac{2}{Re} \left(\alpha^2 (u^*u+v^* v)+Du^*Du+Dv^*Dv\right)&&\nonumber \\
+\int^{(2)} -\frac{2m}{Re} \left(\alpha^2 (u^*u+v^*v)+Du^*Du+Dv^*Dv\right)&&\nonumber \\
+\int_B \Delta \left[u^*\tau_{12}+u\tau^*_{12}+v^*\tau_{22}+v\tau^*_{22}\right] &&\nonumber \\
\end{eqnarray}
where $\Delta f=f^{(2)}-f^{(1)}$ denotes the jump of $f$ over the interface.
Let us develop the remaining boundary term further.
We define $N^{(2)}=N$ and $n^{(2)}=n$ in what follows. The kinematic and dynamic interfacial conditions are (cmp. to \cite{StableWe} where the domains are interchanged): 
\begin{equation}{\Delta u=-h\Delta U', \quad \Delta v=0}\end{equation}
 \begin{eqnarray}
&&N_j\Delta \tau_{ij}+n_j \Delta \mathcal{T}_{ij}+N_j h\Delta\partiel{\mathcal{T}_{ij}}{y}\nonumber \\
&=&We^{-1}\left(\partiel{N_j}{x_j}n_i+\partiel{n_j}{x_j}N_i\right)
\end{eqnarray}
For the Couette problem, the 2nd (dynamic) condition simplifies to:
 \begin{equation}{\Delta \tau_{12}=0, \quad \Delta \tau_{22}=We^{-1}\partiel{n_j}{x_j}}\end{equation}
As $\textbf{n}=(-\mathbf{i}\alpha h,0)$, and $\Delta U'=(1-m)/m$, we obtain:
\begin{eqnarray}
&&\Delta \left[u^*\tau_{12}+v^*\tau_{22}+u \tau_{12}^*+v\tau_{22}^*\right] \nonumber \\
&=&\left[\tau_{12}\Delta u^*+\tau_{12}^* \Delta u\right] +\frac{1}{2}\left[v^*\Delta \tau_{22}+v \Delta \tau_{22}^* \right]\nonumber \\
&=&\frac{1}{Re}\left[(\textrm{i}\alpha v +Du^{(1)})\Delta u^*+(\textrm{i}\alpha v +Du^{(1)})^* \Delta u\right] \nonumber \\
&& -\frac{\alpha^2}{We}\left(v^* h+v h^*\right) \nonumber \\
&=&\frac{(m-1)}{mRe}\left[(\textrm{i}\alpha v +Du)h^*+(-\textrm{i}\alpha v^* +Du^*)h \right] \nonumber \\
&&-\frac{\alpha^2}{We}\left(v^* h+v h^*\right) \nonumber \\
&=&\frac{(m-1)}{mRe}\left[(\textrm{i}\alpha (vh^*-v^*h)+(h^*Du+h Du^*) \right] \nonumber \\
&&-\frac{\alpha^2}{We}\left(v^* h+v h^*\right) \nonumber \\
\end{eqnarray}
and similarly for the complex conjugate of this expression.
Further, we can make use of the interface kinetic condition:
\begin{equation}{\partiel{h}{t}+\textbf{i}\alpha U h=v, \label{eq:kinBC}}\end{equation}
which gives: 
\begin{equation}{\left(v^*h+h^* v\right)=\left(h^*\partiel{h}{t}+h \partiel{h^*}{t}\right)=\partiel{\left(h^*h\right)}{t},}\end{equation}
 and 
\begin{equation}{\left(v^*h-h^* v\right)=\left(h^*\partiel{h}{t}-h \partiel{h^*}{t}\right)=2\textrm{i}{\alpha}{U}h^* h}\end{equation}

\paragraph {Summary} We have arrived at the final energy equation:
\begin{eqnarray}
\frac{d E_{kin}}{dt}&=& \nonumber \\
\int^{(1),(2)} \partiel{||\textbf{u}||^2}{t}&=& \nonumber \\
\int^{(1),(2)} -\left(u^*v+v^*u\right)U'&&\nonumber \\
+\int^{(1)} -\frac{2}{Re} \left(||\alpha \textbf{u}||^2 +||D\textbf{u}||^2\right)&&\nonumber \\
+\int^{(2)} -\frac{2m}{Re} \left(||\alpha \textbf{u}||^2 +||D\textbf{u}||^2\right)&&\nonumber \\
+ W_{We} + W_{m} &&
\end{eqnarray}
where the work performed by the surface tension is:
\begin{equation}{W_{We}=\int_B -\frac{\alpha^2}{We}\partiel{||h||^2}{t}}\end{equation}
and the work performed by viscosity contrast at the interface is:
\begin{equation}{W_{m}=\int_B \frac{(m-1)}{m Re}\left[2\alpha^2 U ||h||^2+Du^{(1)}h^*+Du^{(1)*} h \right]}\end{equation}
The growth rate of the eigenmode can be recovered from $E_{kin}^{-1}(dE_{kin}/dt)$, and hence the components of this expression can be used to analyse how different mechanisms contribute to the eigenvalue growth (the sign tells us whether it is stabilizing or destabilizing, and their relative magnitudes can be compared as well).

The work performed by viscosity contrast may be both positive or negative, depending on the sign of viscosity contrast ($m<1$ or $m>1$) and the relation between $Du^{(1)}$ and $h$.
The work performed by surface tension has a negative sign; as is known, surface tension in itself is purely dissipative in a local temporal sense. However, in the paper it is shown that the surface tension invokes streamwise and vertical velocities near the surface, and when these two are appropriate phase difference they will extract energy from the mean flow shear even without inflection points, similarly to Tollmien-Schlichting waves in boundary layers. 
 
\section{Table summarizing simulation parameters \label{app:tab}}

The following meshes have been used in the nonlinear simulations, all having the same grid spacing in the 
\begin{itemize} 
\item Mesh M1 has 1024000 grid points, domain length $L=250$, grid spacing $0.0078$ times the channel diameter, resulting in xx degrees of freedom.
\item Mesh M2 has 1966080 grid points, domain length $L=120$, grid spacing $0.0039$ times the channel diameter, resulting in xx degrees of freedom.
\item Mesh M3 has 4096000 grid points, domain length $L=250$, grid spacing $0.0039$ times the channel diameter, resulting in xx degrees of freedom.
\end{itemize}

 \begin{table}{
\centering
{\begin{tabular}{|c|c|c|c| l l |c|c| ll | c |c|}
\hline
 $m$  & $Re_{inner}$ & Grid & $L$ & $T_{DNS}$ & $(+T_{sfd})$ & $\Delta T$ & From steady? & Final residual & (averaging time) & Final state\\
\hline
$0.1$ & $40$ & M2 & $120$ & $1189$ &(+626) & 1e-3 & Yes & 5.2e-6 & (100) & Asymmetric steady\\
\hline
$0.1$ & $50$ & M2 & $120$ & 2259 &(+621) & 1e-3 & Yes & 9.8e-4 &(100) & Unsteady\\
\hline
$0.1$ & $60$ & M2 & $120$ & 1305 & & 1e-3 & No & 6e-3 &(100) & Unsteady\\
\hline
$0.1$ & $70$ & M2 & $120$ & 842 & & 1e-3 & No & 7.6e-3& (100) & Unsteady\\
\hline
$0.1$ & $80$ & M2 & $120$ & 814 & & 1e-3 & No & 2.7e-2 & (93)& Unsteady\\
\hline
$0.1$ & $90$ & M2 & $120$ & 1568 & & 1e-3 & No & 3.7e-2 & (90) & Unsteady \\
\hline
$0.1$ & $100$ & M2 & $120$ & 821& & 1e-3 & No & 4.5e-2 &(100) & Unsteady\\
\hline
$0.1$ & $200$ & M2 & $120$ & 498 && 1e-3 & No & 1e-1 & (94) & Unsteady\\
\hline
$0.1667$ & $25$ & M1 & $250$ & 2890 & & 5e-3 & No & 2.0e-8  & (100) & Symmetric steady\\
\hline
$0.1667$ & $25$ & M1 & $250$ & 1979 & (+686) & 2.5e-3 & Yes & 2.0e-8 ($A$=4e-10) & (100)  & Symmetric steady\\
\hline
$0.1667$ & $25$ & M1 & $250$ & 2366  & (+3206)  & 2.5e-3 & Yes ($Re=37.5$) & 2.1e-6  ($A$=3e-6) &(100) & Asymmetric steady\\
\hline
$0.1667$ & $37.5$ & M1 & $250$ & 2555 &(+1263) & 2.5e-3 & Yes & 5.0e-5 ($A$=9e-4) &(100) & Asymmetric steady\\
\hline
$0.1667$ & $37.5$ & M1 & $250$ & 3425& & 5e-3 & No &2.0e-7  &(100) & Asymmetric steady\\
\hline
$0.1667$ & $45.8$ & M1 & $250$ & 2000 &(+1000)& 5e-3 & Yes & 2.0e-8 & (100) & Asymmetric steady\\
\hline
$0.1667$ & $53.3$ & M1 & $250$ & 2000 &(+2000)& 5e-3 & Yes  & 1.8e-8 & (100) & Asymmetric steady\\
\hline
$0.1667$ & $66.7$ & M1 & $250$ & 3904 &(+2000) & 5e-3 & Yes & 1.5e-5 ($A$=3e-8) & (100) & Asymmetric steady\\
\hline
$0.1667$ & $83.3$ & M1 & $250$ & 4712 & (+2000) & 5e-3 & Yes & 3.3e-4 & (100)& Unsteady\\
\hline
$0.2$ & $20$ & M1 & $250$ & 501 && 1e-3 & Yes ($Re=36$) & 2.3e-8 ($A$=3e-11) & (100)& Symmetric steady\\
\hline
$0.2$ & $30$ & M1 & $250$ & 891 && 1e-3 & No & 2.1e-8 & (100) & Symmetric steady\\
\hline
$0.2$ & $30$ & M1 & $250$ & 1247  & (+500) & 2.5e-3 & Yes & 1.9e-8 ($A$=1e-13) & (100) & Symmetric steady\\
\hline
$0.2$ & $30$ & M1 & $250$ & 1014 & (+2639)& 2.5e-3 & Yes ($Re=36$) & 2.1e-7 ($A$=3e-5) & (100) & Asymmetric steady\\
\hline
$0.2$ & $36$ & M1 & $250$ & 2639 & (+4642)& 2.5e-3 & Yes ($Re=40$) & 1.5e-5 ($A$=2e-7) & (100) & Asymmetric steady\\
\hline
$0.2$ & $40$ & M1 & $250$ & 3632 & (+1000)& 2.5e-3 & Yes & 1.1e-5 & (100) & Asymmetric steady\\
\hline
$0.2$ & $50$ & M1 & $250$ & 3405 & (+1000)& 2.5e-3 & Yes &2.0e-6 & (100) & Asymmetric steady\\
\hline
$0.2$ & $80$ & M1 & $250$ & 3205 & (+1281) & 2.5e-3 & Yes & 2.9e-8 & (100) & Asymmetric steady\\
\hline
$0.2$ & $100$ & M1 & $250$ & 5005 & (+660)& 2.5e-3 & Yes & 2.2e-5 & (100) & Unsteady\\
\hline
$0.2$ & $110$ & M1 & $250$ & 5657 && 5e-3 & No & 8.3e-3 & (100) & Unsteady\\
\hline
$0.2$ & $120$ & M1 & $250$ & 2000 & & 5e-3& No & 1.4e-2 & (100) & Unsteady\\
\hline
$0.2$ & $140$ & M1 & $250$ & 1998 && 5e-3 & No & 2.0e-2 & (100) & Unsteady\\
\hline
$0.2$ & $160$ & M1 & $250$ & 1000 && 5e-3 & No & 2.1e-2 & (100) & Unsteady\\
\hline
$0.2857$ & $28.7$ & M1 & $250$ & 2869  & (+832) & 2.5e-3 & Yes & 2.1e-8 ($A$=5e-10) & (100) & Symmetric steady\\
\hline
$0.2857$ & $28.7$ & M1 & $250$ & 878  & (+7000) & 5e-3 & Yes ($Re=42.9$)& 2.1e-8 ($A$=7e-12) & (100) & Symmetric steady\\
\hline
$0.2857$ & $42.9$ & M1 & $250$ & 3897 &  & 5e-3 & No & 9.3e-6 & (100) & Asymmetric steady\\
\hline
$0.2857$ & $42.9$ & M1 & $250$ &  6000 & (+1000)  & 5e-3 & Yes & 9.3e-6 ($A$=3e-7) & (100) & Asymmetric steady\\
\hline
$0.2857$ & $85.7$ & M1 & $250$ & 3925 & (+2000) & 5e-3 & Yes & 1.4e-5 ($A$=1e-8) & (100) & Asymmetric steady\\
\hline
$0.2857$ & $114.3$ & M1 & $250$ & 5342 & (+2000) & 5e-3 & Yes & 1.0e-5 ($A$=4e-9)& (100) & Asymmetric steady\\
\hline
$0.2857$ & $121.4$ & M1 & $250$ & 7325 & (+1000) & 5e-3 & Yes & 1.7e-7 & (100) & Asymmetric steady\\
\hline
$0.2857$ & $128.6$ & M1 & $250$ & 6800 & (+1000) & 5e-3 & Yes & 5.0e-6 ($A$=4e-5)& (100) & Asymmetric steady\\
\hline
$0.2857$ & $142.9$ & M1 & $250$ & 5004 & (+1000) & 5e-3 & Yes & 4.4e-3 ($A$=4e-2) & (100)& Unsteady\\
\hline
$0.3333$ & $50$ & M1 & $250$ & 1675&  & 5e-3 & No & 2.2e-8 ($A$=3e-8) & (100) & Symmetric steady\\
\hline
$0.3333$ & $50$ & M1 & $250$ & 3877  & (+1328) & 5e-3 & Yes & 2.2e-8 & (100) & Symmetric steady\\
\hline
$0.3333$ & $40$ & M1 & $250$ & 1602  & (+3949)  & 5e-3 & Yes ($Re=60$) &2.3e-8 ($A$=2e-8) & (100) &Symmetric steady\\
\hline
$0.3333$ & $50$ & M1 & $250$ & 2377  & (+3949)  & 5e-3 & Yes ($Re=60$) & 7.1e-7 ($A$=4e-4) & (100) &Asymmetric steady\\
\hline
$0.3333$ & $60$ & M1 & $250$ & 2626  & (+1323) & 5e-3 & Yes & 1.4e-6 & (100) & Asymmetric steady\\
\hline
$0.3333$ & $60$ & M1 & $250$ & 3676 & & 5e-3 & No & 2.2e-8 & (100) & Asymmetric steady\\
\hline
$0.3333$ & $66.7$ & M1 & $250$ & 5930 &  & 5e-3 & No & 5.2e-5 & (100) & Asymmetric steady\\
\hline
$0.3333$ & $100$ & M1 & $250$ & 3344 &  & 5e-3 & No & 1.9e-8 & (100) & Asymmetric steady\\
\hline
$0.3333$ & $133.3$ & M1 & $250$ & 1252 & (+1260) & 5e-3 & Yes & 3.0e-6 & (100) & Asymmetric steady\\
\hline
$0.3333$ & $150$ & M1 & $250$ & 2503 & (+1856) & 5e-3 &Yes & 3.9e-3 & (100) & Unsteady\\
\hline
$0.3333$ & $150$ & M1 & $250$ & 4552 &  & 5e-3 & No & 6e-3 & (100) & Unsteady\\
\hline 
$0.3333$ & $166.7$ & M1 & $250$ & 9557 &   & 5e-3 & No & 7.6e-3  & (100) & Unsteady\\
\hline
$0.3333$ & $183.3$ & M1 & $250$ & 3591 &  & 5e-3 & No & 1.6e-2 & (100) & Unsteady\\
\hline
\hline
       \end{tabular}}
       \caption{Simulation parameters, $m=0.1$--$0.3333$, $We=10$.}
   \label{table:t1}
}\end{table}    

 \begin{table}{
\centering
{\begin{tabular}{|c|c|c|c| l l |c|c| ll | c |c|}
\hline
 $m$  & $Re_{inner}$ & Grid & $L$ & $T_{DNS}$ & $(+T_{sfd})$ & $\Delta T$ & From steady? & Final residual &(averaging time) & Final state\\
\hline
$0.4348$ & $54.3$ & M1 & $250$ & 2505  & (+1135) & 5e-3 & Yes ($Re=76.0$) & 2.2e-8 ($A$=3e-9) & (100) & Symmetric steady\\
\hline
$0.4348$ & $65.2$ & M1 & $250$ & 1992 &  & 5e-3 & No & 2.2e-8 & (100) & Symmetric steady\\
\hline
$0.4348$ & $65.2$ & M1 & $250$ & 4000 & (+8341) & 5e-3 & Yes ($Re=76.1$) & 1.2e-7 ($A$=4e-5) & (100) & Asymmetric steady\\
\hline
$0.4348$ & $76.1$ & M1 & $250$ & 1740 & (+6601) & 5e-3 & Yes ($Re=97.8$) & 8.1e-8 ($A$=3e-5) & (100) & Asymmetric steady\\
\hline
$0.4348$ & $87.0$ & M1 & $250$ & 3519 &  & 5e-3 & No &2.1e-8  & (100) & Symmetric steady\\
\hline
$0.4348$ & $87.0$ & M1 & $250$ & 4000 & (+2001) & 5e-3 & Yes & 1.6e-6 ($A$=1e-6) & (100) & Asymmetric steady\\
\hline
$0.4348$ & $97.8$ & M1 & $250$ & 4000 & (+2201) & 5e-3 & Yes & 2.1e-8 & (100) & Asymmetric steady\\
\hline
$0.4348$ & $97.8$ & M1 & $250$ & 5753 &  & 5e-3 & No & 1.2e-5 & (100) & Asymmetric steady\\
\hline
$0.4348$ & $113.9$ & M1 & $250$ & 3420 &  & 5e-3 & No &2.0e-8  & (100) & Asymmetric steady\\
\hline
$0.4348$ & $130.4$ & M1 & $250$ & 4000 & (+1052)  & 5e-3 & Yes & 1.9e-8 & (100) & Asymmetric steady\\
\hline
$0.4348$ & $152.2$ & M1 & $250$ & 5855 & (+2093) & 5e-3 & Yes & 1.8e-8 & (100) & Asymmetric steady\\
\hline
$0.4348$ & $173.9$ & M1 & $250$ & 8949 & (+2090) & 5e-3 & Yes & 2.0e-8 & (100) & Asymmetric steady\\
\hline
$0.4348$ & $195.7$ & M1 & $250$ & 7873 & (+3155) & 5e-3 & Yes & 1.3e-4 & (100) & Unsteady\\
\hline
$0.4348$ & $239.1$ & M1 & $250$ & 5282 & (+3100) & 5e-3 & Yes & 1.6e-2 & (100) & Unsteady\\
 \hline
$0.5$ & $62.5$ & M1 & $250$ & 2613 & (+1200) & 5e-3 & Yes ($Re=80$) & 2.1e-8  & (100) & Symmetric steady \\
 \hline
$0.5$ & $62.5$ & M1 & $250$ & 1983 &  & 5e-3 & No  & 2.2e-8 & (100) & Symmetric steady \\
\hline
$0.5$ & $80$ & M1 & $250$ & 2245 & (+8000) & 5e-3 & Yes ($Re=110$)& 1.0e-6  & (100) & Asymmetric steady \\
\hline
$0.5$ & $95$ & M1 & $250$ & 3737 &  & 5e-3 & No  & 2.2e-8 & (100) & Symmetric steady\\
\hline
$0.5$ & $95$ & M1 & $250$ & 4000 & (+2000) & 5e-3 & Yes & 2.1e-8  & (100) & Asymmetric steady \\
\hline
$0.5$ & $110$ & M1 & $250$ & 6000  & (+2000)  & 5e-3 & Yes  & 1.4e-6 ($A$=2e-6)& (100)  & Asymmetric steady \\
\hline
$0.5$ & $110$ & M1 & $250$ & 6000 &  & 5e-3 & No  & 2.0e-8 & (100) & Asymmetric steady \\
\hline
$0.5$ & $125$ & M1 & $250$ & 6000 & (+1906) & 5e-3 & Yes  & 1.2e-5 ($A$=6e-6) & (100) & Asymmetric steady \\
\hline
$0.5$ & $150$ & M1 & $250$ & 2000 & (+2000) & 5e-3 & Yes  & 1.8e-4 & (100) & Asymmetric steady \\
\hline
$0.5$ & $175$ & M1 & $250$ & 7555 & (+2000)  & 5e-3 & Yes  & 1.8e-8 & (100) & Asymmetric steady \\
\hline
$0.5$ & $200$ & M1 & $250$ & 7156 & (+2000) & 5e-3 & Yes  & 2.0e-8 & (100) & Asymmetric steady \\
\hline
$0.5$ & $212.5$ & M1 & $250$ & 9153  & (+2000) & 5e-3 & Yes  & 1.5e-4 & (100) & Unsteady\\
\hline
$0.5$ & $225$ & M1 & $250$ & 12102  & (+2000)  & 5e-3 & Yes  & 1.3e-4 & (100) & Unsteady \\
\hline
$0.5$ & $250$ & M1 & $250$ & 8000 &  & 5e-3 & No  & 3.0e-4  & (100) & Unsteady\\
\hline
$0.5$ & $262.5$ & M1 & $250$ & 5000  &  & 5e-3 & No  & 4.1e-4 & (100) & Unsteady \\
\hline
$0.5$ & $262.5$ & M1 & $250$ &  1401 & (+3180) & 5e-3 & Yes  & 6.5e-3 & (100) & Unsteady \\
\hline
$0.5$ & $275$ & M1 & $250$ & 2651 & (+4000)  & 5e-3 & Yes  & 2.2e-2 & (100) & Unsteady \\
\hline
$0.6667$ & $83.3$ & M1 & $250$ & 2000 &   & 5e-3 & No   & 2.2e-8  & (100) & Symmetric steady \\
\hline
$0.6667$ & $120$ & M1 & $250$ & 2028 & (+1205)  & 5e-3 & Yes ($Re=146.7$) & 1.4e-7 ($A$=8e-5)  & (100)  & Asymmetric steady\\
\hline
$0.6667$ & $126.7$ & M1 & $250$ & 1787  &   & 5e-3 & No   & 2.1e-8  & (100) & Symmetric steady  \\
\hline
$0.6667$ & $146.7$ & M1 & $250$ & 1630 &   & 5e-3 & No   & 2.1e-8  & (100) & Symmetric steady  \\
\hline
$0.6667$ & $146.7$ & M1 & $250$ & 2000  & (+1168)  & 5e-3 & Yes & 2.1e-8 ($A$=9e-10)  & (100)  & Symmetric steady\\
\hline
$0.6667$ & $146.7$ & M1 & $250$ & 4129  & (+13035)  & 5e-3 & Yes ($Re=166.7$) & 1.1e-6 ($A$=7e-4)  & (100)  & Asymmetric steady\\
\hline
$0.6667$ & $166.7$ & M1 & $250$ & 12000  & (+1035)   & 5e-3 & Yes  & 3.6e-6  & (100) & Asymmetric steady \\
\hline
$0.6667$ & $200$ & M1 & $250$ & 11950  & (+2094)   & 5e-3 & Yes  & 8.9e-6 ($A$=5e-7) & (100) & Asymmetric steady \\
\hline
$0.6667$ & $216.7$ & M1 & $250$ & 5677  & (+1058)   & 5e-3 & Yes   & 7.7e-4 ($A$=3e-4) & (100) &  Unsteady\\
\hline
$0.6667$ & $233.3$ & M1 & $250$ & 5977  & (+1011)   & 5e-3 &  Yes  & 4.4e-4  & (100) &  Unsteady\\
\hline
$0.6667$ & $283.3$ & M1 & $250$ & 3537  & (+4315)  & 5e-3 & Yes   & 7.0e-3  & (100) &  Unsteady \\
\hline
$0.6667$ & $316.7$ & M1 & $250$ & 3120  & (+3180)   & 5e-3 & Yes   & 2.1e-2  & (100) & Unsteady \\
\hline
$0.7692$ & $173.1$ & M1 & $250$ & 2657 & (+2000) & 5e-3 & Yes ($Re=211.5$) & 2.0e-8  ($A$=4e-7)& (100)  & Symmetric steady \\
\hline
$0.7692$ & $192.3$ & M1 & $250$ & 1601 &   & 5e-3 & No  & 1.9e-8  & (100)  & Symmetric steady  \\
\hline
$0.7692$ & $192.3$ & M1 & $250$ & 2669  & (+2000) & 5e-3 & Yes ($Re=211.5$) &1.9e-8 ($A$=2e-6) & (100)  & Symmetric steady  \\
\hline
$0.7692$ & $211.5$ & M1 & $250$ & 3934 &   & 5e-3 &  No & 1.9e-8 & (100)  & Symmetric steady  \\
\hline
$0.7692$ & $211.5$ & M1 & $250$ & 4000 & (+1162)  & 5e-3 &  Yes & 1.9e-8 ($A$=7e-10) & (100)  & Symmetric steady \\
\hline
$0.7692$ & $211.5$ & M1 & $250$ &  & (+)  & 5e-3 &  Yes ($Re=230.8$) &  & (100)  & Symmetric steady \\
\hline
$0.7692$ & $230.8$ & M1 & $250$ & 4000  & (+1009)   & 5e-3 & Yes  & 2.3e-8  & (100)  & Asymmetric steady  \\
\hline
$0.7692$ & $250$ & M1 & $250$ & 11618  & (+1135)  & 5e-3 & Yes   & 6.8e-4 & (100)  &  Unsteady \\
\hline
$0.7692$ & $269.2$ & M1 & $250$ & 8000  & (+3178)   & 5e-3 & Yes   & 2.6e-4  & (100)  &  Unsteady \\
\hline
$0.7692$ & $296.2$ & M1 & $250$ & 6423  & (+1078)   & 5e-3 & Yes   & 2.7e-2  & (100)  & Unsteady  \\
\hline
$0.7692$ & $326.9$ & M1 & $250$ & 5279  & (+3171)   & 5e-3 & Yes   & 3.2e-2  & (100)  & Unsteady  \\
\hline
$0.9091$ & $254.5$ & M1 & $250$ & 6000 & (+6591)   & 5e-3 & Yes   & 1e-5 ($A$=3e-7) & (100) & Steady  \\
\hline
$0.9091$ & $272.7$ & M1 & $250$ & 6000  & (+2906)   & 5e-3 & Yes   & 2.8e-4 ($A$=1e-3) & (100) & Unsteady  \\
\hline
$1$ & $230$ & M1 & $250$ & & - & - & No & - && Symmetric steady \\
\hline
$1$ & $245$ & M1 & $250$ & 2000 & (+) & - & Yes & 1e-5 ($A$=5e-11)&& Symmetric steady\\
\hline
$1$ & $250$ & M1 & $250$ && - & - & No & - && Unsteady\\
\hline
$1$ & $500$ & M1 & $250$ && - & - & No & - && Unsteady\\
\hline
 \end{tabular}}
       \caption{Simulation parameters, $m=0.4348$--$1$, $We=10$.}
   \label{table:t2}
}\end{table}    

\begin{table}{
\centering
{\begin{tabular}{|c|c|c|c| l l |c|c| ll | c |c|}
\hline
 $m$  & $Re_{inner}$ & Grid & $L$ & $T_{DNS}$ & $(+T_{sfd})$ & $\Delta T$ & From steady? & Final residual &(average time) & Final state\\
\hline
$1.1$ & $230$ & M1 & $250$ & 1983   & (+1250)   & 5e-3 & Yes   & 3.4e-8 ($A$=1.1e-10) & (100) & Steady \\
\hline
$1.1$ & $250$ & M1 & $250$ & 2000  & (+4059)   & 5e-3 & Yes   & 1.9e-8 ($A$=1.2e-11) & (100) & Steady  \\
\hline
$1.1$ & $270$ & M1 & $250$ & 3174  & (+4308)   & 5e-3 & Yes   & 4e-3 ($A$=1.4e-2) & (100) & Unsteady  \\
\hline
$1.5152$ & $150$ & M1 & $250$ & 1257  & (+1932)   & 5e-3 & Yes   & 2.2e-8   & (100) & Symmetric steady \\
\hline
$1.5152$ & $225$ & M1 & $250$ & 2606  &  (+2177) & 5e-3 & Yes   & 2.0e-8  & (100) & Symmetric steady  \\
\hline
$1.5152$ & $275$ & M1 & $250$ & 5128  & (+1156)  & 5e-3 & Yes   & 1.1e-6  ($A$=1e-7) & (100) & Symmetric steady  \\
\hline
$1.5152$ & $310$ & M1 & $250$ & 4423  & (+1230)  & 5e-3 & Yes   & 3.5e-6 ($A$=4e-7) & (100) & Symmetric steady  \\
\hline
$1.5152$ & $350$ & M1 & $250$ & 5359   &  (+1207) & 5e-3 & Yes   & 4.6e-4 ($A$=7e-4) & (100) & Unsteady  \\
\hline
$2$ & $90$ & M1 & $250$ & 1295   &  (+2000) & 5e-3 & Yes   & 2.2e-8   & (100) & Symmetric steady  \\
\hline
$2$ & $120$ & M1 & $250$ & 2611   &  (+2042) & 5e-3 & Yes   & 2.2e-8  & (100) & Symmetric steady  \\
\hline
$2$ & $150$ & M1 & $250$ & 1286   &  (+1167) & 5e-3 & Yes   & 2.2e-8  & (100) & Symmetric steady  \\
\hline
$2$ & $200$ & M1 & $250$ & 3785   &  (+2248) & 5e-3 & Yes   & 4.9e-6 ($A$=9e-8)  & (100) & Symmetric steady  \\
\hline
$2$ & $250$ & M1 & $250$ & 4521   &  (+1182) & 5e-3 & Yes   & 2.5e-6 ($A$=4e-7)  & (100) & Symmetric steady \\
\hline
$2$ & $300$ & M1 & $250$ & 4973   &  (+1113) & 5e-3 & Yes   & 1.3e-2 ($A$=2e-3)  & (100) & Unsteady  \\
\hline
$2$ & $350$ & M1 & $250$ &  1589  &  (+1086) & 5e-3 & Yes   & 1.9e-2 ($A$=3e-2) & (100) & Unsteady  \\
\hline
$3$ & $200$ & M1 & $250$ & 4790 &  & 5e-3 & No   &1.6e-6  & (100) & Symmetric steady  \\
\hline
$3$ & $200$ & M1 & $250$ & 3949  &  (+1315) & 5e-3 & Yes   & 4.6e-7 ($A$=2e-6) & (100) & Symmetric steady  \\
\hline
$3$ & $225$ & M1 & $250$ & 4646  &  (+1943) & 5e-3 & Yes   & 1.1e-4 ($A$=1e-4) & (100) & Unsteady  \\
\hline
$3$ & $250$ & M1 & $250$ & 1705  &  & 5e-3 & No   & 1.2e-2  & (100) & Unsteady  \\
\hline
$3$ & $300$ & M1 & $250$ & 2442  &  & 5e-3 & No   & 2.1e-2  & (100) & Unsteady  \\
\hline
$3$ & $400$ & M1 & $250$ & 721 &  & 5e-3 & No   & 2.5e-2  & (100) & Unsteady  \\
\hline
$3$ & $500$ & M1 & $250$ & 787 &  & 5e-3 & No   & 3.3e-2  & (100) & Unsteady  \\
\hline
$3$ & $600$ & M1 & $250$ & 757 &  & 5e-3 & No   & 4.5e-2  & (100) & Unsteady  \\
\hline
$3$ & $700$ & M1 & $250$ & 775 &  & 5e-3 & No   & 5.0e-2  & (100) & Unsteady  \\
  \hline
$5$ & $50$ & M1 & $250$ & 795  & (+689)  & 5e-3 & Yes   & 5.7e-7 ($A$=2e-6) & (100) & Symmetric steady \\
 \hline
$5$ & $100$ & M1 & $250$ & 2095  & (+1110)  & 1e-3 & Yes   & 5.8e-6 ($A$=3e-4) & (100) & Symmetric steady\\
\hline
$5$ & $200$ & M1 & $250$ & 1463  & (+442) & 1e-3 & Yes   & 1.5e-3 ($A$=2e-2) & (100) & Unsteady  \\
\hline
$5$ & $200$ & M1 & $250$ & 1148  &  & 1e-3 & No   & 1.8e-3 & (100) & Unsteady  \\
\hline
$5$ & $300$ & M1 & $250$ & 3178  &  & 5e-3 & No   & 2.9e-2 & (100) & Unsteady  \\
\hline
$5$ & $500$ & M1 & $250$ & 786  &  & 5e-3 & No   & 5.2e-2 & (100) & Unsteady  \\
\hline
$5$ & $600$ & M1 & $250$ & 799  &  & 5e-3 & No   & 7.3e-2 & (100) & Unsteady  \\
\hline
$5$ & $700$ & M1 & $250$ & 728  &  & 5e-3 & No   & 9.4e-2 & (100) & Unsteady  \\
\hline
$5$ & $800$ & M1 & $250$ & 803  &  & 5e-3 & No   & 1.1e-1 & (100) & Unsteady  \\
\hline
$5$ & $900$ & M1 & $250$ & 783  &  & 5e-3 & No   & 1.1e-1 & (100) & Unsteady  \\
\hline
$10$ & $75$ & M1 & $250$ & 1464  &  & 5e-3 & No & 2.2e-2 & (100) & Unsteady  \\ 
\hline
$10$ & $150$ & M1 & $250$ & 1077  & (+800) & 5e-3 & Yes & 3.6e-2 & (100) & Unsteady  \\
\hline
$10$ & $250$ & M1 & $250$ & 1111  & (+800)  & 5e-3 & Yes & 5.2e-2 & (100) & Unsteady  \\
\hline
$10$ & $300$ & M1 & $250$ & 1098  & (+793) & 5e-3 & Yes & 1.7e-1 & (100) & Unsteady  \\
\hline
 \end{tabular}}
       \caption{Simulation parameters, $m=1.1$--$10$, $We=10$.}
   \label{table:t3}
}\end{table}    

\begin{table}{
\centering
{\begin{tabular}{|c|c|c|c| l l |c|c| ll | c |c|}
\hline
 $m$  & $Re_{inner}$ & Grid & $L$ & $T_{DNS}$ & $(+T_{prev.})$ & $\Delta T$ & Initial condition? & Final residual &(average time) & Final state\\
    \hline
$0.2$ & $20$ & M1 & $250$ & 634 & (+1014)& 2.5e-3 & Yes ($We=10$, $Re=30$, asymm.) & 6.6e-5 ($A$=4e-5)  & (100) & Steady \\
   \hline
$0.2$ & $30$ & M1 & $250$ & 1657  & (+1014)& 2.5e-3 & Yes ($We=10$, asymm.) & 2.8e-3 ($A$=3e-3)  & (100) & Unsteady\\
  \hline
$0.5$ & $45$ & M1 & $250$ & 1982 & (+2613)& 5e-3 & Yes ($We=10$, $Re=62.5$, symm.) & 2.3e-8 ($A$=9e-12) & (100) & Symmetric steady\\
  \hline
$0.5$ & $62.5$ & M1 & $250$ & 4362  & (+6000)& 5e-3 & Yes ($We=10$, $Re=110$, symm.) & 5.9e-4 & (100) & Unsteady\\
 \hline
$0.5$ & $80$ & M1 & $250$ & 3237  & (+2245)& 5e-3 & Yes ($We=10$, asymm.) & 3.1e-3 & (100) & Unsteady\\
 \hline
$0.5$ & $95$ & M1 & $250$ & 3497  & (+4000)& 5e-3 & Yes ($We=10$, symm.) & 3.6e-3  & (100) & Unsteady \\
 \hline
$2$ & $250$ & M1 & $250$ & 2428   & (+4521)& 5e-3 & Yes ($We=10$, symm.) & 5.6e-5 & (100) & Symmetric steady\\
\hline
$2$ & $300$ & M1 & $250$ & 3877  & (+4973)& 5e-3 & Yes ($We=10$, symm.) & 1.9e-8 & (100) & Symmetric steady \\
\hline
$2$ & $350$ & M1 & $250$ & 3279  & (+1589)& 5e-3 & Yes ($We=10$, unsteady) & 9e-7 & (100) & Symmetric steady \\
\hline
$2$ & $400$ & M1 & $250$ & 4318  & (+3279)& 5e-3 & Yes ($Re=350$) & 1.7e-6 & (100) & Symmetric steady \\
\hline
$2$ & $450$ & M1 & $250$ & 1653 & (+4973) & 5e-3 & Yes ($We=10$, unsteady) & 7.2e-5  & (100) &  Symmetric steady \\
\hline
$10$ & $250$ & M1 & $250$ & 618  & (+4000)& 1e-3 & Yes ($We=10$, unsteady) & 8.7e-4 ($A$=6e-5) & (100) & Steady\\
\hline
$10$ & $300$ & M1 & $250$ & 693  & (+4000)& 1e-3 & Yes ($We=10$, unsteady) & 2.0e-2 ($A$=9e-5) & (100) & Unsteady (insp.) \\
\hline
 \end{tabular}}
       \caption{Simulation parameters, $We=2$.}
   \label{table:t5}
}\end{table}

\begin{table}{
\centering
{\begin{tabular}{|c|c|c|c| l l |c|c| ll | c |c|}
\hline
 $m$  & $Re_{inner}$ & Grid & $L$ & $T_{DNS}$ & $(+T_{prev.})$ & $\Delta T$ & Initial condition? & Final residual &(average time) & Final state\\
 \hline
$0.2$ & $20$ & M1 & $250$ & 2697  & (+1159)& 1e-3 & Yes ($Re=30$) & 5.0e-5 ($A$=4e-10) & (100) & Symmetric steady\\
\hline
$0.2$ & $30$ & M1 & $250$ & 1159 & (+3653)& 2.5e-3 & Yes ($We=10$) & 2.7e-8 ($A$=3e-8) & (100) & Asymmetric steady\\
\hline
$0.2$ & $80$ & M1 & $250$ & 1888 & (+1492)& 2.5e-3 & Yes & 9.6e-3 ($A$=2e-1) & (100) & Unsteady \\
\hline
$0.2$ & $90$ & M1 & $250$ & 1700 & (+925)& 2.5e-3 & Yes & 1.1e-2 ($A$=3e-1) & (100) & Unsteady \\
\hline
$0.5$ & $50$ & M1 & $250$ & 3160 & (+2050)& 2.5e-3 & Yes ($Re=160$) & 4.3e-5  & (100) & Steady  \\
\hline
$0.5$ & $80$ & M1 & $250$ & 2050 & (+2245)& 2.5e-3 & Yes ($We=10$) & 2.7e-3  & (100) & Unsteady  \\
\hline
$10$ & $150$ & M1 & $250$ & 1059 & (+768)& 1e-3 & Yes & 1.2e-5 ($A$=1e-11) & (100) & Symmetric steady\\
\hline
$10$ & $300$ & M1 & $250$ & 1051 & (+1098)& 1e-3 & Yes & 1.5e-1 ($A$=7e-5) & (100) & Unsteady\\
\hline
 \end{tabular}}
       \caption{Simulation parameters, $We=5$.}
   \label{table:t5}
}\end{table}    

\begin{table}{
\centering
{\begin{tabular}{|c|c|c|c| l l |c|c| ll | c |c|}
\hline
 $m$  & $Re_{inner}$ & Grid & $L$ & $T_{DNS}$ & $(+T_{prev.})$ & $\Delta T$ & Initial condition? & Final residual &(average time) & Final state\\
 \hline
$0.2$ & $20$ & M1 & $250$ & 356 & (+1256)& 1e-3 & Yes ($Re=30$) & 3e-6 ($A$=1e-9) & (100) & Symmetric steady\\
\hline
$0.2$ & $30$ & M1 & $250$ & 1256 & (+3653)& 2.5e-3 & Yes ($We=10$) & 1.9e-8 ($A$=4e-9) & (100) & Asymmetric steady\\
\hline
$0.2$ & $80$ & M1 & $250$ & 5133 & (+1460) & 2.5e-3 & Yes & 1.4e-4 ($A$=4e-8) & (100) & Asymmetric steady \\
\hline
$0.2$ & $90$ & M1 & $250$ & 3204 & (+1219) & 2.5e-3 & Yes & 1.8e-3 ($A$=1e-3)  & (100) & Unsteady \\
\hline
$2$ & $300$ & M1 & $250$ & 3883 & (+4973) & 5e-3 & Yes ($We=10$, unsteady) & 5.2e-6 ($A$=5e-9) & (100) & Symmetric steady \\
\hline
$2$ & $250$ & M1 & $250$ & 2193 & (+4973) & 5e-3 & Yes ($We=10$, steady) & 6.3e-7 & (100) & Symmetric steady \\
\hline
 \end{tabular}}
       \caption{Simulation parameters, $We=20$.}
   \label{table:t5}
}\end{table}    

\begin{table}{
\centering
{\begin{tabular}{|c|c|c|c| l l |c|c| ll | c |c|}
\hline
 $m$  & $Re_{inner}$ & Grid & $L$ & $T_{DNS}$ & $(+T_{prev.})$ & $\Delta T$ & Initial condition? & Final residual &(average time) & Final state\\
  \hline
$0.5$ & $200$ & M1 & $250$ & 1457 & (+3781) & 5e-3 & Yes ($We=10$, unsteady) & 6.0e-3 ($A$=3e-7) & (100) & Unsteady  (insp.)\\
 \hline
$0.5$ & $250$ & M1 & $250$ & 2140 & (+4973) & 5e-3 & Yes ($We=10$, unsteady) & 6.7e-3 ($A$=7e-7) & (100) & Unsteady (insp.)  \\
\hline
$0.5$ & $300$ & M1 & $250$ & 1467 & (+4973) & 5e-3 & Yes ($We=10$, unsteady) & 7.5e-3 ($A$=7e-7) & (100) & Unsteady (insp.) \\
\hline
 \end{tabular}}
       \caption{Simulation parameters, $We=\infty$.}
   \label{table:t5}
}\end{table}    

\end{document}